# Photonuclear reactions on stable isotopes of cadmium and tellurium at bremsstrahlung end-point energies of 10-23 MeV

F.A. Rasulova[a,b,*], A.A. Kuznetsov[c,d], V.O. Nesterenko[e], J.H. Khushvaktov[b,f], S.I. Alekseev[a], N.Yu. Fursova[c], A.S. Madumarov[a], I. Chuprakov[a,g], S.S. Belyshev[c,d], N.V. Aksenov[a]

[a] Flerov Laboratory of Nuclear Reactions of Joint Institute for Nuclear Research, Dubna, 141980 Russia
[b] Institute of Nuclear Physics of the Academy of Sciences of the Republic of Uzbekistan, Tashkent, 100214 Republic of Uzbekistan
[c] Skobeltsyn Institute of Nuclear Physics of Lomonosov Moscow State University, Moscow, 119234 Russia
[d] Faculty of Physics of Lomonosov Moscow State University, Moscow, 119991 Russia
[e] Bogoliubov Laboratory of Theoretical Physics of Joint Institute for Nuclear Research, Dubna, Russia
[f] Dzhelepov Laboratory of Nuclear Problems of Joint Institute for Nuclear Research, Dubna, 141980 Russia
[g] Institute of Nuclear Physics, Almaty, 050032 Republic of Kazakhstan

*rasulova@jinr.ru

**Abstract**

This work used the γ-activation approach to conduct tests at bremsstrahlung end-point energies of 10-23 MeV utilising the MT-25 microtron beam. The experimental values of relative yields and cross sections per equivalent quantum of photonuclear reactions on stable isotopes of cadmium and tellurium were compared to theoretical calculations obtained from TALYS-2.0 using the default parameters and a combined model of photonucleon reactions (CMPR). The inclusion of isospin splitting in the combined model of photonucleon reactions allows for the description of experimental data on proton escape reactions with energies ranging from 17 to 23 MeV. As a result, isospin splitting must be taken into consideration in order to accurately describe the decay of the giant dipole resonance. For Cd isotopes, essential discrepancies of yet unclear origin between theory (TALYS 2.0 and CMPR) and experimental data are found in the neutron channel.

## 1. Introduction

In order to learn more about how the population of nuclear shells affects unique aspects of the decay of excited nuclear states, it is interesting to investigate the photodisintegration of nuclei located close to the $Z = 50$ closed proton shell. The yields of different channels of the photodisintegration of isotopes compared to the ratio of the numbers of neutrons and protons in nuclei can highlight the mechanisms of excitation and decay of nuclear states in the energy region between 10 and 50 MeV.

Cadmium ($Z = 48$) and tellurium ($Z = 52$) isotopes are convenient objects of investigation, since under natural conditions, there exist several stable isotopes, which allows to obtain the dependence of the yields of various reactions on the number of neutrons in a nucleus. The main decay channels of giant dipole resonance (GDR) are the emission of neutrons or protons. Currently, there is an extensive database of experimental data on photoneutron reactions in the stable isotopes of natural mixture of cadmium and tellurium [1-22]. Nevertheless, these data are not yet complete and properly explained. The proton channel, despite a small cross section of the ($\gamma,p$) reaction, is interesting in connection with isospin splitting of GDR [23].

Natural cadmium consists of eight stable isotopes with following mass numbers and isotopic abundances: $^{106}$Cd (1.25%), $^{108}$Cd (0.89%), $^{110}$Cd (12.49%), $^{111}$Cd (12.80%), $^{112}$Cd (24.13%), $^{113}$Cd (12.22%), $^{114}$Cd (28.73%) and $^{116}$Cd (7.49%). The photoneutron reaction cross sections $\sigma(\gamma,n) + \sigma(\gamma, np)$ and $\sigma(\gamma, 2n)$ as well as the total absorption cross section $\sigma(\gamma, sn) = \sigma(\gamma,n) + \sigma(\gamma, np) + \sigma(\gamma, 2n)$ for a target from a natural mixture of cadmium isotopes were measured by using a beam of quasimonochromatic photons without separating the contributions from the reactions on individual isotopes [1]. Absolute majority of photonuclear reaction data on cadmium isotopes has been obtained in experiments using bremsstrahlung photons, namely, relative yields of multiparticle reactions on natural cadmium were studied using end-point energy of 23 MeV [2,3], 55 MeV [4-6], the flux-averaged cross section were determined using end-point energies of 50 and 60 MeV [7], isomeric ratios were determined using electron bremsstrahlung for pairs $^{115m,g}$Cd [7-15] and $^{104m,g}$Ag [13].

Natural tellurium consists of eight stable isotopes with following mass numbers and isotopic abundances: $^{120}$Te (0.09%), $^{122}$Te (2.55%), $^{123}$Te (0.89%), $^{124}$Te (4.74%), $^{125}$Te (7.07%), $^{126}$Te (18.84%), $^{128}$Te (31.74%) and $^{130}$Te (34.08%). Although all of these isotopes can undergo photodisintegration via different reaction channels, only a small number have been studied so far, namely, photoneutron reactions [10-12,16-22]. The cross sections for the ($\gamma,n$) [22], ($\gamma,n$) + ($\gamma,pn$) [16] and ($\gamma,2n$) + ($\gamma,2np$) [16] reactions on the isotopes $^{120,124,126,128,130}$Te induced by bremsstrahlung photons and positron annihilation in flight were determined by detecting neutrons in the energy range of $\gamma$ quantum 8.03-26.46 MeV. Isomeric ratios have been measured for the pairs $^{119m,g}$Te [10,17,19,20], $^{121m,g}$Te [10,11,19,20], $^{123m,g}$Te [10,18], $^{127m,g}$Te [11,19,21] and $^{129m,g}$Te [10-12,17,19,21].

This work aims to obtain new data on fundamental photonuclear reactions on cadmium and tellurium isotopes using a bremsstrahlung $\gamma$-radiation beam with energies between 10 and 23 MeV. We used TALYS-2.0 [24] and combined model of photonucleon reactions (CMPR) [25] for simulations and the $\gamma$-activation method with bremsstrahlung photons from the electron accelerator to obtain our experimental data. Furthermore, the photoproton reaction product the $^{111}$Ag is prospective medicinal isotope [26-29]; thus, examining the reaction yields is a beneficial for both research and application.

The paper is organized as follows. In Sec. 2 the experimental set-up and procedures are described. In Sec. 3 the methods of data analysis are outlined. In Sec. 4, the results for Cd and Te isotopes are presented and discussed. In Sec. 5, the conclusions are drawn. In Appendix 1, the tabulated experimental results are presented. In Appendix 2, The TALYS options and GDR isospin splitting are inspected.

## 2. Experimental Set-up and procedures

This experiment was carried out using the MT-25 microtron's output electron beam [30]. The electron energies ranged from 10 to 23 MeV, with an energy step of 1 MeV. Table 1 lists the main parameters of the experiments. To generate γ-radiation, a tungsten radiator target, a common convertor material, was employed. The tungsten target was thick enough (3 mm) to maximise the amount of photons in the energy range of the giant dipole resonance (GDR), which dominates the photonuclear cross section between the nucleon separation threshold and 20-30 MeV. To eliminate the leftover electrons from the bremsstrahlung beam, a 30 mm thick aluminium absorber was positioned behind the tungsten converter. The Cd and Te targets were located perpendicular to the electron beam and 1 cm away from the converter. The target of natural cadmium had dimensions 10×10×0.5 mm (at 10–19 MeV) and 5×5×0.5 mm (at 20–23 MeV). The natural tellurium target in a form of a powder in an aluminium foil in the form of a square envelope with a size of 15×15×2 mm (at 10-15 MeV), 8×8×2 mm (at 16-19 MeV) and 6×6×2 mm (at 20-23 MeV).

In the trials, a bremsstrahlung flux generated in the tungsten converter was utilised to irradiate metallic natural cadmium and tellurium samples. The fluctuations in beam current were measured using a Faraday cup and a calibrated ionisation chamber in the beam, and then recorded into a web-accessible database for use in the study with LabView software and an analog-to-digital converter card. [31]. Along with the Faraday cup and ionisation chamber, the beam current was determined by digitising the electrical charge accumulated on the target. During irradiation, the electron current of the accelerator was measured using a Faraday cup. A 0.15-mm-thick copper monitor was positioned behind the irradiation target. The absolute value of the current was calculated by comparison of experimentally measured and theoretical yields at the monitor on the basis of the $^{65}Cu(\gamma,n)^{64}Cu$ reaction. The experimental cross sections of the partial photoneutron reactions for the $^{65}Cu$ nucleus, obtained on quasimonoenergetic annihilation photon beams [32] using the neutron multiplicity sorting method display considerable systematic uncertainties and do not satisfy the specially introduced objective physical data reliability criteria. We used the corrected theoretical cross sections which were used to evaluate the cross sections of the partial reactions using their experimental-theoretical approach [33]. The yield of the $^{65}Cu(\gamma,n)^{64}Cu$ reaction was determined using the expected cross section [33], and the bremsstrahlung spectrum was generated using Geant4 [34].

Once the radiation levels in the experimental hall were safe, the targets were transferred to a different measurement room and the induced activity in the irradiated target was measured. We used a high purity germanium (HPGe) γ-detector with a resolution of 16 keV at 1332 keV, together with standard measurement electronics and a 16K ADC/MCA (Multiport II Multichannel Analyser, CANBERRA). The energy and efficiency of the HPGe detector were calibrated using standard γ-ray sources. A thorough explanation of the γ-activation measurement method used in this study can be found in [35,36].

Table 1. Main parameters of the experiments

| Energy of electrons (MeV) | Mass of targets (mg) | | Integral charge (mC) | | Irradiation time (min) | | Total measuring time of spectra (days) | |
|---|---|---|---|---|---|---|---|---|
| | $^{nat}$Cd | $^{nat}$Te | $^{nat}$Cd | $^{nat}$Te | $^{nat}$Cd | $^{nat}$Te | $^{nat}$Cd | $^{nat}$Te |
| 10 | 475 | 1050 | 50 (5) | 30 (3) | 97 | 100 | 1.8 | 3.9 |
| 11 | 448 | 1100 | 50 (5) | 20 (2) | 150 | 62 | 1.8 | 3.1 |
| 12 | 454 | 1010 | 50 (5) | 10 (1) | 125 | 32 | 1.8 | 3.8 |
| 13 | 451 | 1060 | 50 (5) | 10 (1) | 155 | 29 | 2.0 | 3.2 |
| 14 | 451 | 1080 | 30 (3) | 4.0 (4) | 175 | 50 | 1.6 | 3.6 |
| 15 | 414 | 1000 | 20 (2) | 4.0 (4) | 65 | 12 | 1.7 | 3.5 |
| 16 | 417 | 220 | 10 (1) | 4.0 (4) | 65 | 15 | 1.7 | 3.8 |
| 17 | 427 | 190 | 4.0 (4) | 4.0 (4) | 37 | 34 | 1.6 | 4.7 |
| 18 | 401 | 100 | 3.0 (3) | 4.0 (4) | 24 | 49 | 1.5 | 4.7 |
| 19 | 389 | 110 | 3.0 (3) | 4.0 (4) | 32 | 55 | 1.6 | 3.5 |
| 20 | 113 | 90 | 3.0 (3) | 4.0 (4) | 51 | 28 | 1.6 | 2.9 |
| 21 | 116 | 90 | 3.0 (3) | 4.0 (4) | 35 | 41 | 1.5 | 3.3 |
| 22 | 116 | 90 | 3.0 (3) | 4.0 (4) | 27 | 37 | 1.7 | 4.7 |
| 23 | 112 | 90 | 3.0 (3) | 4.0 (4) | 21 | 31 | 1.7 | 4.1 |

The duration that elapsed between the end of the irradiation process and the beginning of the measurement was between 10 and 15 minutes, designated as the cooling period. The spectra of each sample were taken many times over a total of 0.5, 1, 12, and 24 hours. Fig. 1 and Fig. 2 display typical γ-ray spectra of the chemical products generated from the $^{nat}$Cd and $^{nat}$Te, respectively. A background spectrum (red line) is also shown in Fig. 1. Bremsstrahlung radiation with end-point energy of 23 MeV was used to irradiate the samples.

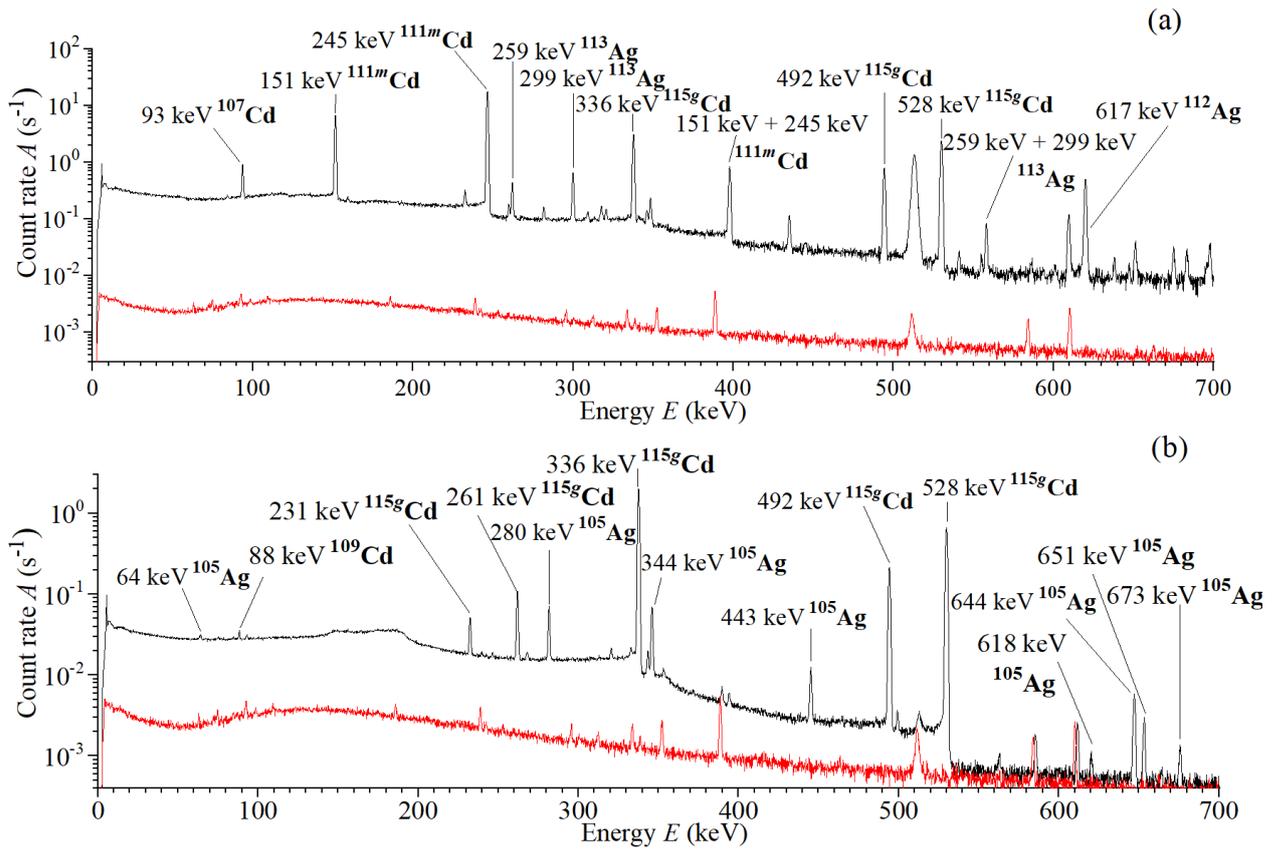

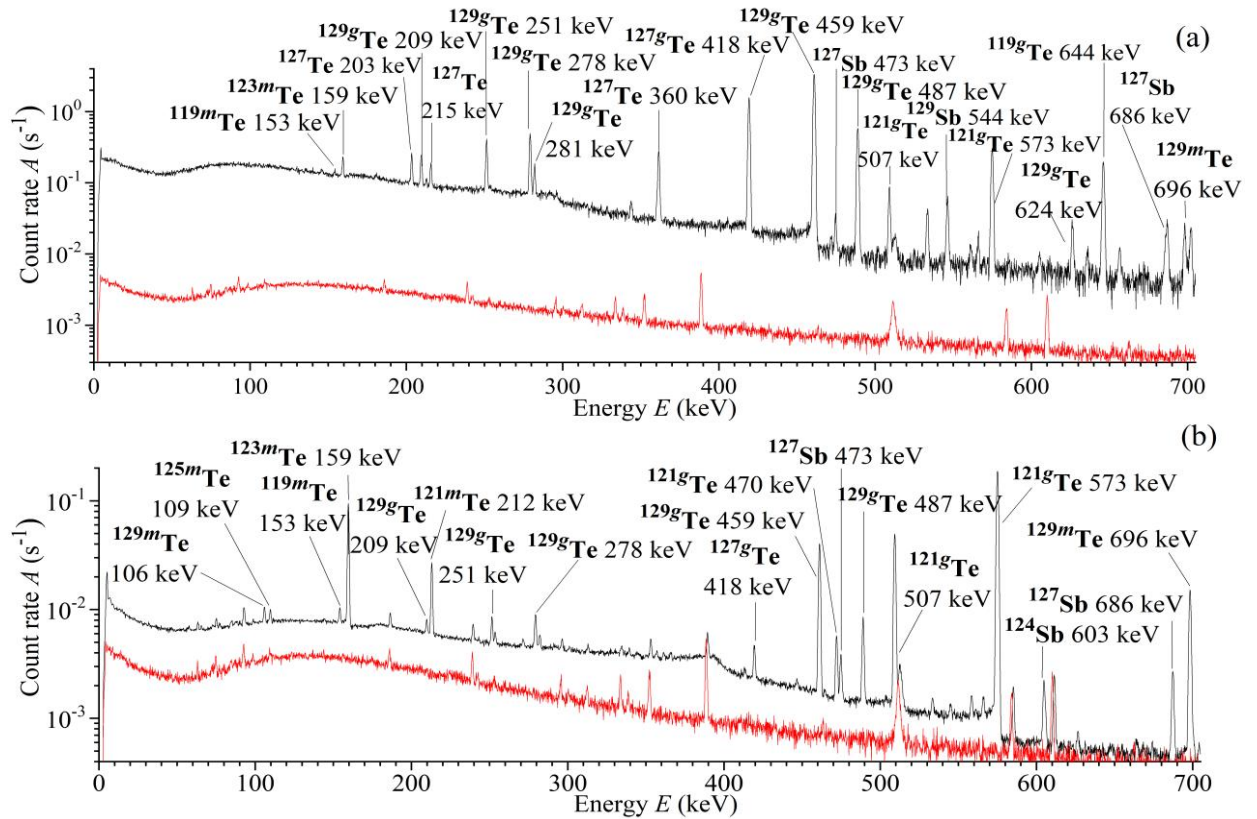

Figure 1. Spectra of residual activity of the irradiated sample of $^{nat}$Cd (top-to-bottom) 3 h (*a*) and 4 days (*b*) after irradiation. The spectra measurement duration was 1 h (*a*) and 1 day (*b*), respectively

Figure 2. Spectra of residual activity of the irradiated sample of $^{nat}$Te (top-to-bottom) 6 h (a) and 11 days (b) after irradiation. The spectra measurement duration was 1 h (a) and 3 days (b), respectively

The γ-ray spectra were processed using the DEIMOS32 code [37]. This code uses the Gaussian function to fit the count area of full-energy peaks. The processed peaks were identified using the half-life of residual nuclei, γ-ray energy, and intensity. The radionuclides were identified based on their different γ-ray energies and half-lives. Table 2 provides the key γ-ray energies and intensities used to compute the reaction product yield. Table 2's columns 4-5 contain nuclear data from Ref. [38].

Table 2. Spectroscopic data from ref. [38] for the product nuclei from the photonuclear reactions on stable isotopes of cadmium and tellurium

| Nucleus | Half-life $T_{1/2}$ | Decay mode (%) | γ-ray energy $E_\gamma$ (keV) ($I_\gamma$ (%)) | Reaction | $E_{th}$ (MeV) |
|---|---|---|---|---|---|
| Data for irradiated cadmium target ||||||
| $^{105}$Cd | 55.5 m | EC | 346.87 (4.2), 961.84 (4.69) | $^{106}$Cd(γ,n) | 10.9 |
| $^{107}$Cd | 6.5 h | EC | 93.124 (4.7) | $^{108}$Cd(γ,n) | 10.3 |
| $^{109}$Cd | 461.9 d | EC | 88.03 (3.644) | $^{110}$Cd(γ,n)+ $^{111}$Cd(γ,2n) | 9.9 16.89 |
| $^{111m}$Cd | 48.54 m | IT | 150.82 (29.1), 245.39 (94) | $^{112}$Cd(γ,n)+ $^{113}$Cd(γ,2n) | 9.4 15.93 |
| $^{115g}$Cd | 53.46 h | β− | 336.24 (46.02), 527.9 (27.45) | $^{116}$Cd(γ,n) | 8.7 |
| $^{115m}$Cd | 44.56 d | β− | 933.8 (2) | $^{116}$Cd(γ,n) | 8.9 |
| $^{105}$Ag | 41.29 d | EC | 280.41 (30.2), 344.52 (41) | $^{106}$Cd(γ,p) | 7.4 |
| $^{111}$Ag | 7.45 d | β− | 342.13 (6.7) | $^{112}$Cd(γ,p)+ $^{113}$Cd(γ,np) | 9.6 16.19 |
| $^{112}$Ag | 3.13 h | β− | 617.51 (43), 1387.68 (5.3) | $^{113}$Cd(γ,p)+ $^{114}$Cd(γ,np) | 9.7 18.76 |
| $^{113}$Ag | 5.37 h | β− | 298.6 (10) | $^{114}$Cd(γ,p) | 10.3 |
| $^{115}$Ag | 20 m | β− | 229.1 (18) | $^{116}$Cd(γ,p) | 11 |
| Data for irradiated tellurium target ||||||
| $^{119g}$Te | 16.05 h | EC | 644.01 (84.1), 699.85 (10.1) | $^{120}$Te(γ,n) | 10.29 |
| $^{119m}$Te | 4.7 d | EC | 153.59 (66), 1212.73 (66.1) | $^{120}$Te(γ,n) | 10.55 |
| $^{121g}$Te | 19.17 d | EC | 573.14 (80.4) | $^{122}$Te(γ,n)+ $^{123}$Te(γ,2n) | 9.83 16.76 |
| $^{121m}$Te | 164.2 d | IT: 88.6 EC: 11.4 | 212.19 (81.5) | $^{122}$Te(γ,n)+ $^{123}$Te(γ,2n) | 10.12 17.05 |
| $^{123m}$Te | 119.2 d | IT | 159.0 (84.3) | $^{123}$Te(γ,γ`)+ $^{124}$Te(γ,n)+ $^{125}$Te(γ,2n) | 9.67 16.24 |
| $^{125m}$Te | 57.4 d | IT | 109.28 (0.28) | $^{125}$Te(γ,γ`)+ $^{126}$Te(γ,n) | 9.25 |
| $^{127}$Te | 9.35 h | β− | 417.9 (0.99) | $^{128}$Te(γ,n) | 8.78 |
| $^{129g}$Te | 69.6 min | β− | 459.60 (7.7) | $^{130}$Te(γ,n) | 8.42 |
| $^{129m}$Te | 33.6 d | IT: 64 EC: 36 | 695.88 (3.0) | $^{130}$Te(γ,n) | 8.52 |
| $^{122}$Sb | 2.72 d | β−:97.59 EC: 2.41 | 564.24 (70.67) | $^{123}$Te(γ,p)+ $^{124}$Te(γ,np) | 8.13 17.55 |
| $^{124}$Sb | 60.2 d | β− | 602.72 (97.8), 1690.97 (47.57) | $^{125}$Te(γ,p)+ $^{126}$Te(γ,np) | 8.69 17.80 |
| $^{127}$Sb | 3.85 d | β− | 473.0 (25.8), 685.7 (36.8) | $^{128}$Te(γ,p) | 9.58 |
| $^{129}$Sb | 4.366 h | β− | 812.97 (48.2), 914.96 (23.3) | $^{130}$Te(γ,p) | 10.01 |

The half-lives of previously investigated radionuclides ranged from 20 min ($^{115}$Ag) to 461.9 days ($^{109}$Cd) as well as from 69.6 min ($^{129g}$Te) to 164.2 days ($^{121m}$Te). To compute radioactive half-lives and select appropriate spectra for each isotope's activity, γ-ray spectra were collected throughout a range of waiting times, from minutes to a day after irradiation. The activity is typically measured using the highest intensity, well-separated, interference-free, and correctable γ-ray.

## 3. Data analysis

The experimental yields of the reactions $Y_{exp}$ were normalized to one electron of the accelerated beam incident on the bremsstrahlung target and calculated using the following formula:

$$Y_{exp} = \frac{S_p \cdot C_{abs}}{\varepsilon \cdot I_\gamma} \frac{t_{real}}{t_{live}} \frac{1}{N} \frac{1}{N_e} \frac{e^{\lambda \cdot t_{cool}}}{(1 - e^{-\lambda \cdot t_{real}})} \frac{\lambda \cdot t_{irr}}{(1 - e^{-\lambda \cdot t_{irr}})}, \qquad (1)$$

where $S_p$ is the full-energy-peak area; $\varepsilon$ is the full-energy-peak detector efficiency; $I_\gamma$ is the γ- emission probability; $C_{abs}$ is the correction for self-absorption of γ-rays in the sample; $t_{real}$ and $t_{live}$ are the real time and live time of the measurement, respectively; $N$ is the number of atoms in the activation sample; $N_e$ is the integral number of incident electrons; $\lambda$ is the decay constant; $t_{cool}$ is the cooling time; and $t_{irr}$ is the irradiation time.

The experiment determined the yields $Y_{theor}$ of photonuclear reactions, which reflect the convolution of the photonuclear reactions cross section $\sigma_i(E)$, and the distribution density of the number of bremsstrahlung photons over energy per one electron of the accelerator $W(E, E_{\gamma max})$. The outcome of measuring the yield of isotope generation in all possible reactions on a natural mixture of isotopes is as follows:

$$Y_{theor}(E_{\gamma max}) = \sum_i \eta_i \int_{E_{i\,th}}^{E_{\gamma max}} \sigma_i(E) W(E, E_{\gamma max}) dE \qquad (2)$$

where $E_{\gamma max}$ is the kinetic energy of electrons hitting the tungsten radiator, $E$ is the energy of bremsstrahlung photons produced on the radiator, $E_{th}$ is the threshold of the studied photonuclear reaction, $\eta_i$ is the percentage of the studied isotope in the natural mixture, and the index $i$ corresponds to the number of the reaction contributing to the production of the studied isotope.

Figure 3 illustrates the distribution density of the number of bremsstrahlung photons $W(E, E_{\gamma max})$ per one electron of the accelerator for accelerated electron energies from 10 to 23 MeV, determined using Geant4 for the bremsstrahlung target made of tungsten with a thickness of 3 mm.

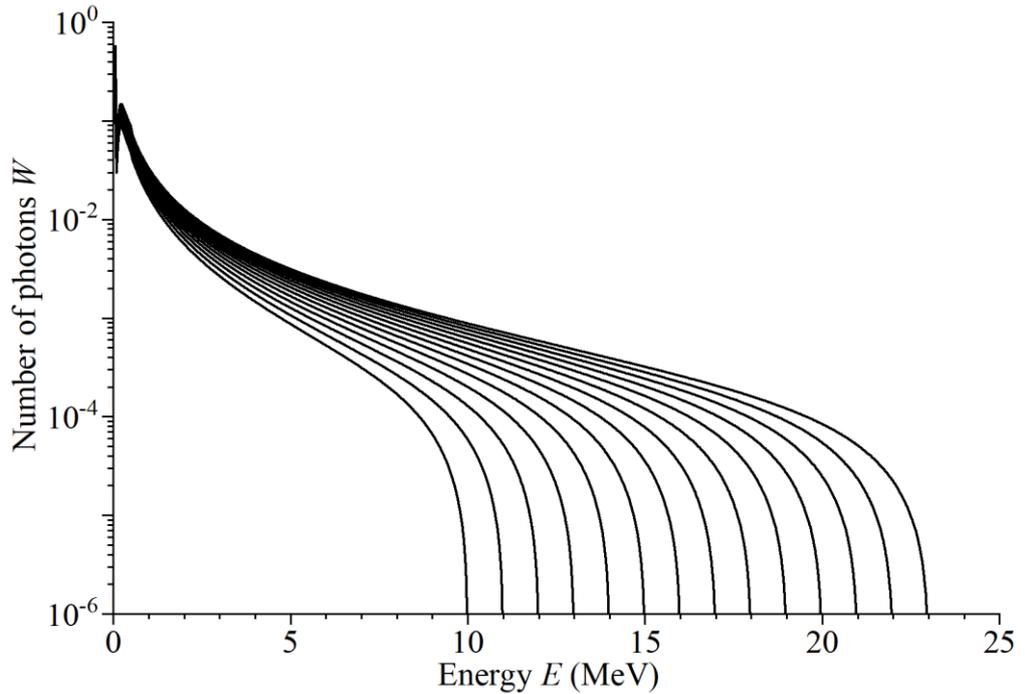

Figure 3. The distribution density of the number of bremsstrahlung photons at the energies of 10-23 MeV

The total and partial cross sections $\sigma(E)$ of photonuclear reactions on cadmium and tellurium isotopes were estimated for monochromatic photons using the TALYS code [24] with standard parameters, and CMPR [25]. The TALYS program examines all processes in the nucleus and transitions between states. As a result, it is possible to calculate not only the total cross sections of photonuclear reactions, but also the cross sections of reactions involving the production of certain states, particularly isomeric states. The standard (default) TALYS option uses Simple Modified Lorentzian (SMLO) model for photon strength functions (PSF). This model is used to calculate E1, M1, and upbend components, generally providing more accurate, temperature-dependent resonance shapes than older models.

As compared with TALYS, the CMPR accurately takes into account the GDR isospin splitting, which is of a crucial importance for description of the proton decay channel. The basics of the GDR isospin splitting and some relevant CMPR results are presented in Appendix 2. This Appendix also provides the information on TALYS options as well as examples of comparison of TALYS and CMPR results

The yield measurement for a natural mixture of isotopes yields gives the amount of isotope produced in all potential reactions on the natural mixture. The primary problem of bremsstrahlung beam experiments is that the yield of photonuclear reaction depends on both the investigated cross section of the reaction $\sigma_i(E)$ and the form of the bremsstrahlung spectrum $W(E, E_{\gamma max})$, which is often known with inadequate accuracy. The use of relative yields allows us to determine the dependency of the yield of photonuclear reactions on the maximal energy of bremsstrahlung under various experimental settings. The overall photon absorption cross section is not taken into account when calibrating the yield of one of the most likely reactions. The most probable and well-studied $^{116}Cd(\gamma,n)^{115}Cd$ and

$^{130}$Te($\gamma$,n)$^{129}$Te reactions were chosen as a primary reaction in case of cadmium and tellurium, respectively. Also there are no other channels (for example, ($\gamma$,2n) reaction on heavier stable nuclei) for product formation of $^{115}$Cd and $^{129}$Te since the nuclei $^{116}$Cd and $^{130}$Te are the heaviest stable nuclei in the natural mixture of Cd and Te, respectively.

Theoretical values of the relative yields can be calculated using the following formula:

$$Y_{rel}(E_{\gamma max}) = \frac{\sum_i \eta_i \int_{E_{th}}^{E_{\gamma max}} \sigma_i(E) W(E, E_{\gamma max}) dE}{\eta \int_{E_{th}}^{E_{\gamma max}} \sigma_{(\gamma,n)}(E) W(E, E_{\gamma max}) dE} . \quad (3)$$

where $\eta$ is the percentage of the $^{116}$Cd and $^{130}$Te isotopes in the natural mixture of cadmium and tellurium isotopes, respectively. Owing to the assumption on the unchanged shape of the bremsstrahlung spectrum, the bremsstrahlung spectrum $W(E, E_{\gamma max})$ can be replaced by the photon production cross section $\sigma(E, E_{\gamma max})$ calculated using Seltzer-Berger tables [39]:

$$Y_{rel}(E_{\gamma max}) = \frac{\sum_i \eta_i \int_{E_{th}}^{E_{\gamma max}} \sigma_i(E) \sigma(E, E_{\gamma max}) dE}{\eta \int_{E_{th}}^{E_{\gamma max}} \sigma_{(\gamma,n)}(E) \sigma(E, E_{\gamma max}) dE} . \quad (4)$$

To represent the experimental photonuclear reaction data, the cross section per equivalent quantum $\sigma_q$ is used determined by the expression:

$$\sigma_q(E_{\gamma max}) = \frac{\int_{E_{th}}^{E_{\gamma max}} \sigma(E) \sigma(E, E_{\gamma max}) dE}{\frac{1}{E_{\gamma max}} \int_0^{E_{\gamma max}} E \cdot \sigma(E, E_{\gamma max}) dE} . \quad (5)$$

The cross section per equivalent quantum for a natural mixture of isotopes includes all possible channels of the final isotope production with account for the percentage of initial nuclei is:

$$\sigma_q^{nat}(E_{\gamma max}) = \frac{\sum_{i=1}^{8} \eta_i \int_{E_{th}}^{E_{\gamma max}} \sigma_i(E) \sigma(E, E_{\gamma max}) dE}{\frac{1}{E_{\gamma max}} \int_0^{E_{\gamma max}} E \cdot \sigma(E, E_{\gamma max}) dE} . \quad (6)$$

The experimental points along the cross sections of the ($\gamma$,n) [22] and ($\gamma$,n) + ($\gamma$,pn) [16] reactions on the isotopes $^{120,128,130}$Te were approximated by the Lorentz function, the relative yields $Y_{rel}$ and cross sections per equivalent quantum $\sigma_q$ were calculated based on the least squares approximation. In Fig. 4-10 these points are indicated by open circles [22] and open rectangles [16], respectively.

## 4. Results and discussion

The measured relative yields to the yield of the reaction $^{116}$Cd($\gamma$,n)$^{115}$Cd (in the case of $^{nat}$Cd) and $^{130}$Te($\gamma$,n)$^{129}$Te (in the case of $^{nat}$Te) and cross sections per equivalent quantum for a natural mixture of isotopes are shown in Fig. 4-25 and Tables 1-2 in Appendix, along with theoretical computations using the TALYS and CMPR programs and previously published data.

The square root of the quadratic sum of all independent statistical and systematic uncertainties was used to determine the overall uncertainties in the results. The counting statistics from the observed number of counts under the photo-peak of each γ-line (2.5%~10.5%) were the primary contributors to the ensuing statistical uncertainty. Data accumulation for an optimal duration of measurements based on the half-life of the generated nuclides was used to estimate this. The systematic uncertainties, on the other hand, were computed using the uncertainties of the following: the number of target nuclei (~0.3%), the irradiation and cooling time (~0.5%), the current and electron beam energy (~2.5%), the detector efficiency (~3%), the half-life of the reaction products (~2%), the distance between the sample and detector (~2%), the γ-ray abundance (~2%), the flux estimation (~11.5%) and normalization of the experimental data to the $^{116}$Cd(γ,n)$^{115}$Cd and $^{130}$Te(γ,n)$^{129}$Te monitor reactions' yield 0.5-2%. Roughly 12.58% is the overall systematic uncertainty. It is determined that the overall uncertainty ranges from about 12% to about 19%.

## A. Photonuclear reactions on cadmium isotopes
### A1. Photoneutron reactions

Five cadmium isotopes are directly generated by $^{nat}$Cd(γ, n) reactions when natural cadmium is irradiated with bremsstrahlung radiation with end-point energy of 10-23 MeV. In this study, the relative yields and cross section per equivalent quantum of the $^{nat}$Cd(γ, n)$^{105,107,109,111m,115g,115m}$Cd reactions at the bremsstrahlung end-point energies of 10-23 MeV are determined and presented in Fig. 4-8. Besides, the tabulated results are given in Appendix 1.

1. $^{106}$Cd(γ, n)$^{105}$Cd reaction

For this reaction, no literature data was available; hence, its measurements were only compared with the theoretical calculations. Fig. 4 shows experimental obtained and simulated values relative yields as well as cross section per equivalent quantum of reaction $^{106}$Cd(γ, n)$^{105}$Cd. In Fig. 4, it is clear that the cross section per equivalent quantum calculated by the TALYS and CMPR codes are almost the same, but they are higher than the currently presented results for the $^{106}$Cd(γ, n)$^{105}$Cd reaction.

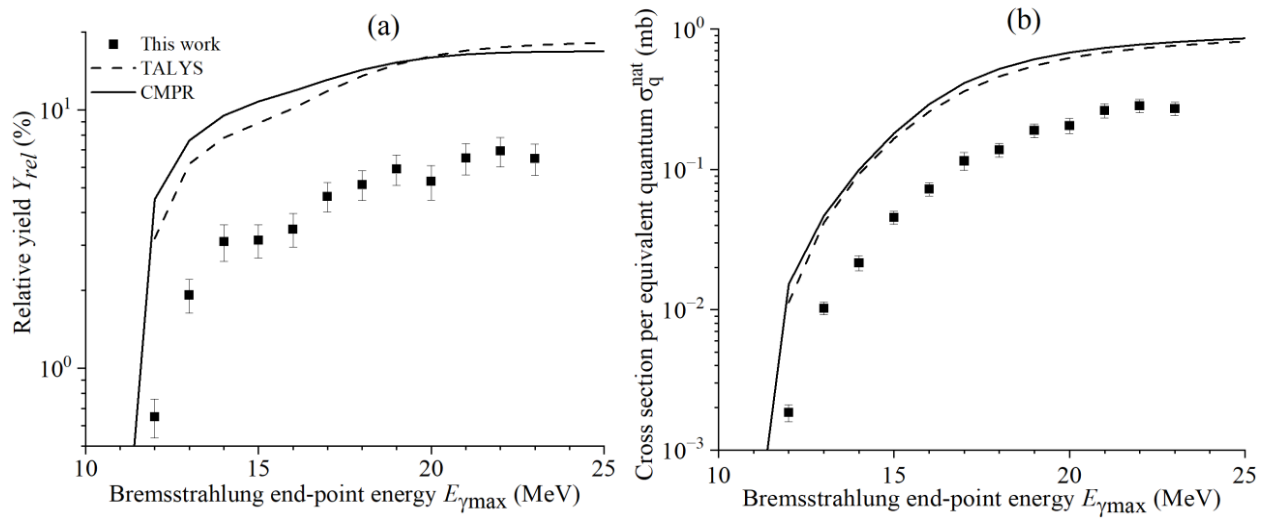

Figure 4. Relative yields (*a*) and cross section per equivalent quantum (*b*) of reaction $^{106}$Cd($\gamma$, $n$)$^{105}$Cd as a function of bremsstrahlung end-point energy from the present work (solid rectangles) and simulated values using the CMPR (solid lines) and TALYS code (dashed lines)

At an energy of 12 MeV, the theoretical values for the two models are 4 times higher than the experimental point, then with an increase in energy to 20 MeV, the ratio $\sigma_{qexp}^{nat}/\sigma_{qtheory}^{nat}$ decreases and is equal to approximately 2.5. The apparent disparity between theory and experiment may stem from the fact that statistical models of photonuclear reactions, such as the CMPR and TALYS codes, overlook the unique structural characteristics of cadmium isotopes. Alternatively, the low experimental neutron yield in reaction $^{106}$Cd($\gamma$, $n$)$^{105}$Cd can be partly explained by bypassed character of the nucleus $^{106}$Cd.

2. $^{108}$Cd($\gamma$, $n$)$^{107}$Cd reaction

Fig. 5 shows that the current results follow the graphical shape but are lower than the theoretical values. The experimental results ($\gamma$,$n$) are two times as less as its theoretical counterparts. This discrepancy indicates the need for further research in the study of photoneutron reactions on $^{106,108}$Cd.

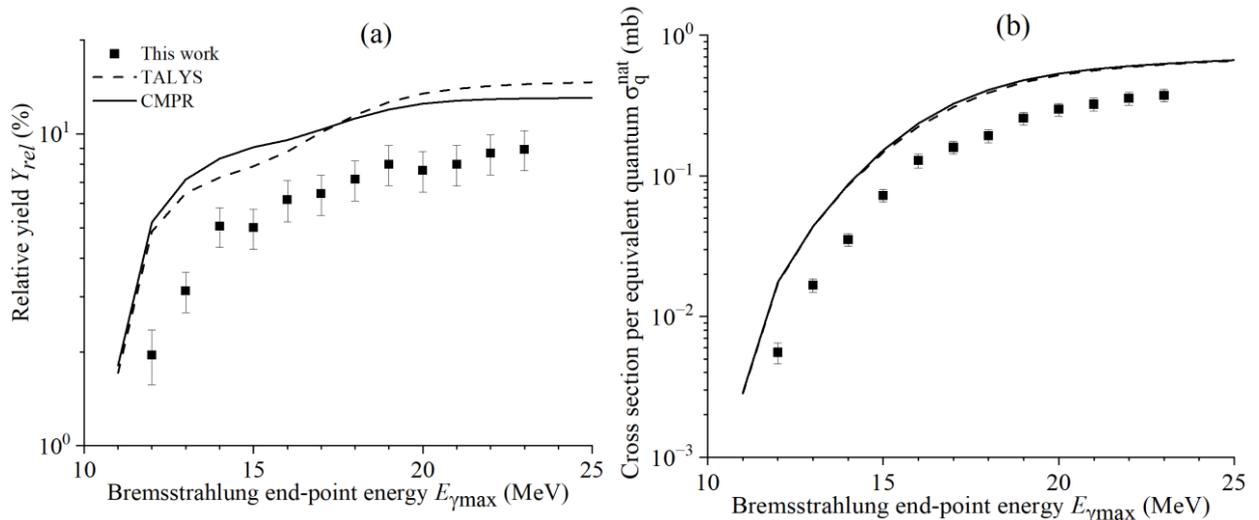

Figure 5. Relative yields (*a*) and cross section per equivalent quantum (*b*) of reaction $^{108}$Cd($\gamma$, $n$)$^{107}$Cd as a function of bremsstrahlung end-point energy from the present work (solid rectangles) as well as simulated values using the CMPR (solid lines) and TALYS code (dashed lines)

3. $^{110}$Cd($\gamma$, $n$)$^{109}$Cd and $^{111}$Cd($\gamma$, $2n$)$^{109}$Cd reactions

For the production of $^{109}$Cd from the $^{nat}$Cd($\gamma$, $Xn$)$^{109}$Cd reactions, only one previous experimental data sets in the GDR energy region based on bremsstrahlung photons was available [3]; this was found to be lower than the theoretical values obtained using the TALYS and CMPR codes as shown in Fig. 6. The figure shows that the current results follow the graphical shape but are lower than the theoretical values; they are the closest to the values calculated using TALYS code.

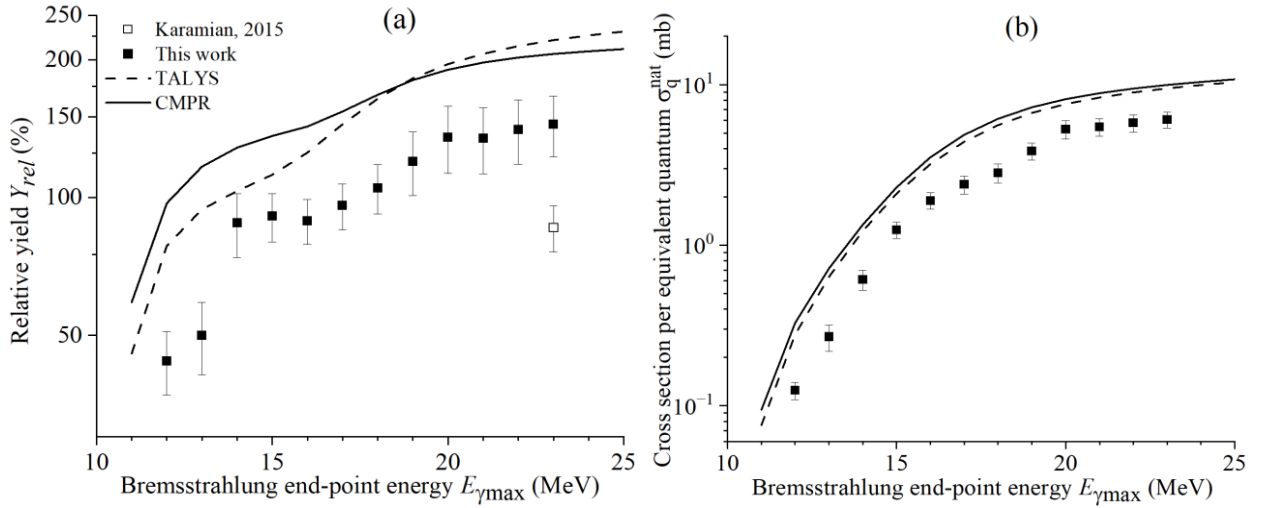

Figure 6. Relative yields (*a*) and cross section per equivalent quantum (*b*) of reactions $^{110}$Cd($\gamma$, $n$)$^{109}$Cd and $^{111}$Cd($\gamma$, $2n$)$^{109}$Cd as a function of bremsstrahlung end-point energy from the present work (solid rectangles), literature data [3] (open rectangle) as well as simulated values using the CMPR (solid lines) and TALYS code (dashed lines)

4. $^{111}$Cd($\gamma$, $\gamma`$)$^{111m}$Cd and $^{112}$Cd($\gamma$, $n$)$^{111m}$Cd reactions

The measured results for the $^{111}$Cd($\gamma$, $\gamma`$)$^{111m}$Cd and $^{112}$Cd($\gamma$, $n$)$^{111m}$Cd reactions are compared with the theoretical values obtained with the TALYS and CMPR codes, as shown in Fig. 7. For the production of $^{111m}$Cd from the $^{nat}$Cd target, only one previous experimental data set in the GDR energy region based on bremsstrahlung photons were available [3], this was found to be lower than the theoretical values obtained using the TALYS code as shown in Fig. 7. The figure also shows that the theoretical values from both the TALYS and CMPR codes as well as literature data are in agreement with the data from this study. Based on the Fig. 7, it can also be said that in the energy range up to 11 MeV, the $^{111m}$Cd nucleus is formed due to the $^{111}$Cd($\gamma$, $\gamma`$)$^{111m}$Cd reaction, the contributions of the $^{111}$Cd($\gamma$, $\gamma`$)$^{111m}$Cd and $^{112}$Cd($\gamma$, $n$)$^{111m}$Cd reactions to the formation of $^{111m}$Cd are equal at 12 MeV, and then the $^{112}$Cd($\gamma$, $n$)$^{111m}$Cd reaction plays a dominant role.

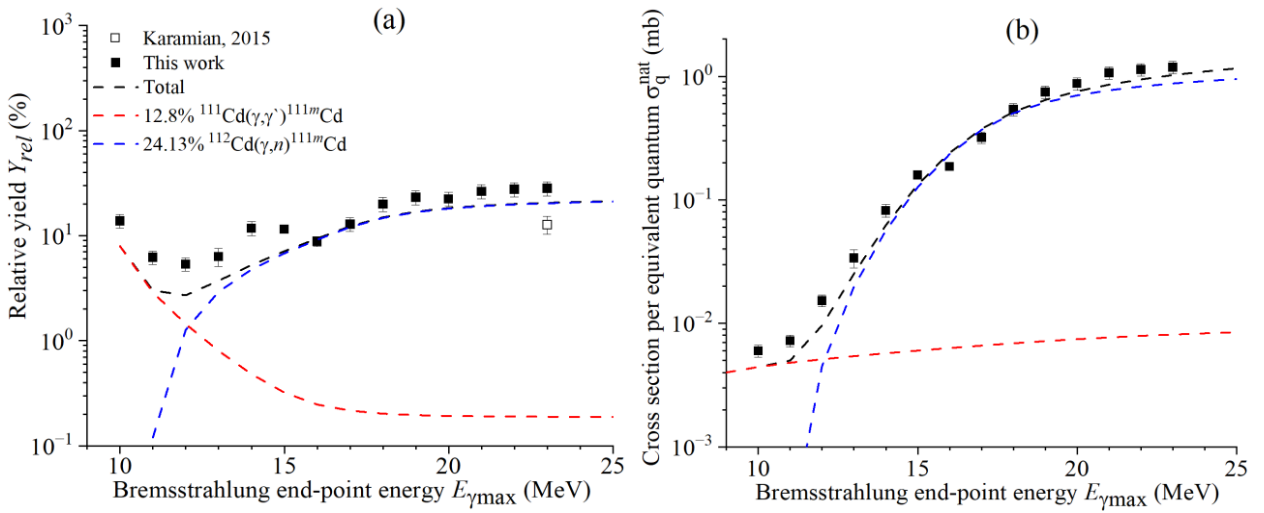

Figure 7. Relative yields (*a*) and cross section per equivalent quantum (*b*) of $^{111}$Cd($\gamma, \gamma'$)$^{111m}$Cd and $^{112}$Cd($\gamma, n$)$^{111m}$Cd reactions as a function of bremsstrahlung end-point energy from the present work (solid rectangles), literature data [3] (open rectangle) and simulated values using TALYS code (dashed lines)

### 5. $^{116}$Cd($\gamma, n$)$^{115m,g}$Cd reaction

The measured results for the $^{116}$Cd($\gamma, n$)$^{115m,g}$Cd reactions are compared with the theoretical values obtained with the TALYS and CMPR codes, as shown in Fig. 8. There is only one literature data for the $^{116}$Cd($\gamma, n$)$^{115m,g}$Cd reaction [3] in 23 MeV. As shown in Fig. 8, it is clear that the theoretical values from both the TALYS and CMPR codes as well as literature data are in agreement with the data from this study.

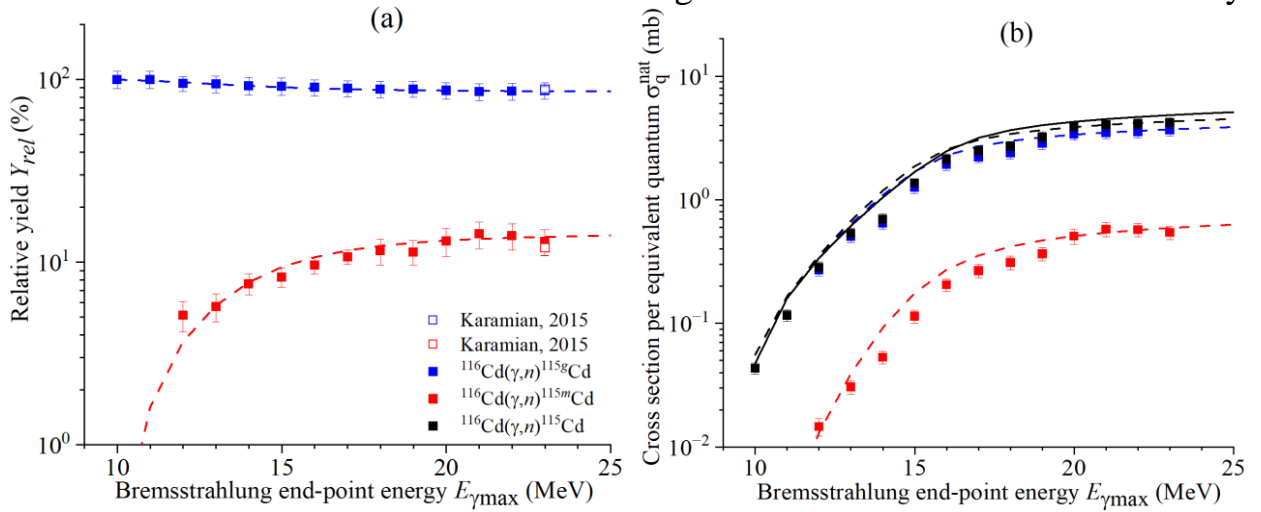

Figure 8. Relative yields (*a*) and cross section per equivalent quantum (*b*) of reaction $^{116}$Cd($\gamma, n$)$^{115}$Cd as a function of bremsstrahlung end-point energy from the present work (solid rectangles), literature data [3] (open rectangles) as well as simulated values using the CMPR (solid line) and TALYS code (dashed lines)

### A2. Photoproton reactions

Five argentum isotopes are directly generated by $^{nat}$Cd($\gamma, p$) reactions when natural cadmium is irradiated with bremsstrahlung radiation with end-point energy of 10-23 MeV. In this study, the relative yields and cross section per equivalent quantum of the $^{nat}$Cd($\gamma, p$)$^{105,111,112,113,115}$Ag reactions at the bremsstrahlung end-point energies of 12-23 MeV are determined for the first time and presented in Fig. 9-14.

### 1. $^{106}$Cd($\gamma, p$)$^{105}$Ag reaction

These measurements were only compared with the theoretical values due to unavailability of published data. In Fig. 9, the measured values are revealed to be higher than the calculated values. No theoretical calculation can describe the experimental data.

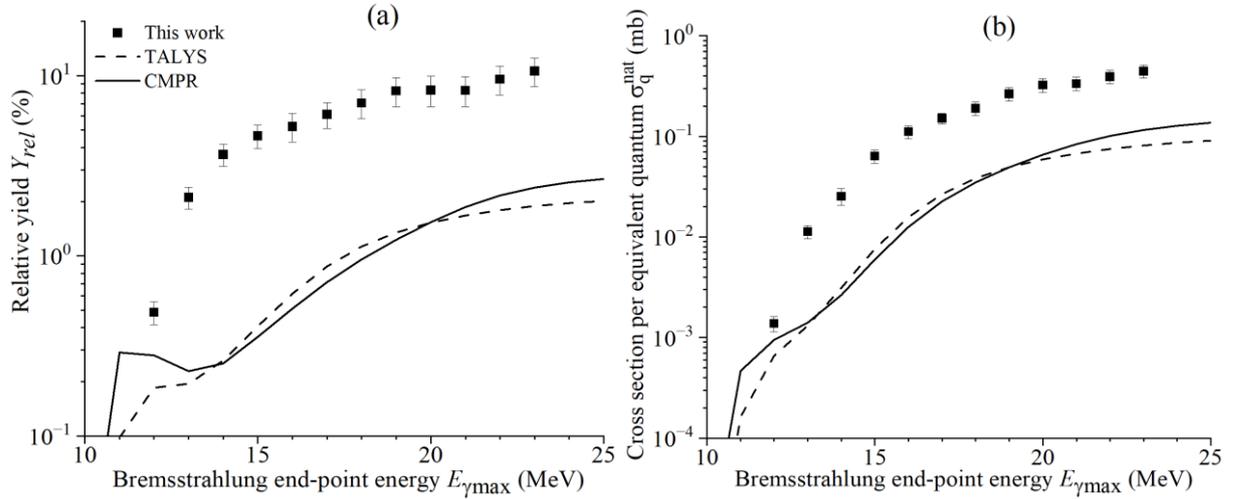

Figure 9. Relative yields (*a*) and cross section per equivalent quantum (*b*) of reaction $^{106}$Cd($\gamma$, p)$^{105}$Ag as a function of bremsstrahlung end-point energy from the present work (solid rectangles) and simulated values using the CMPR (solid lines) and TALYS code (dashed lines)

The ratio $\sigma_{qexp}^{nat}/\sigma_{qtheory}^{nat}$ is approximately 10 for the two models. As we discussed in Section A1 that, the proton Fermi surface is significantly higher than the neutron Fermi surface in the $^{106}$Cd nucleus. As a result of a reduction in the effective width of the Coulomb barrier, this must raise the proton penetrabilities and, consequently, the proton yield.

Proton yield on $^{106}$Cd can also be boosted by the direct photoelectric effect, which is predominantly localised at the nucleus' surface. Protons are expected to dominate in $\beta^+$-radioactive nuclei and nuclei bordering them. The target nucleus's structural (shell) unique features have a significant impact on direct photonuclear reactions. Cadmium isotopes exhibit the most significant single-particle dipole excitations during the $1g_{9/2} \to 1h_{11/2}$ transitions. The decay of such excitations, which have a probability of about 40% in both the proton and neutron channels, leads to the emission of protons with the maximum possible energy, because the final-state nucleus {Z-1, N} arises in the ground state. The energy carried away by neutrons is 3 to 4 MeV smaller, because the final-state nucleus {Z, N-1} remains in an excited (hole) state. This significantly influences the relative output of protons and neutrons in proton-rich cadmium isotopes.

Fig. 10 shows the relative yields of $^{106}$Cd($\gamma$, n)$^{105}$Cd and $^{106}$Cd($\gamma$, p)$^{105}$Ag reactions as well as sum of the $^{106}$Cd($\gamma$, n)$^{105}$Cd and $^{106}$Cd($\gamma$, p)$^{105}$Ag reactions. The results in Fig. 10 show that the theoretical yields of the ($\gamma$,n) and ($\gamma$,p) reactions on $^{106}$Cd nuclei differ significantly from their experimental counterparts, but their summed value agrees with the respective experimental value. This means that the original photoabsorption cross section calculated using the TALYS code is unlikely to differ significantly from the true value. The discrepancy in cross sections per equivalent quantum is due to the redistribution of the cross section between the ($\gamma$,n) and ($\gamma$,p) reactions.

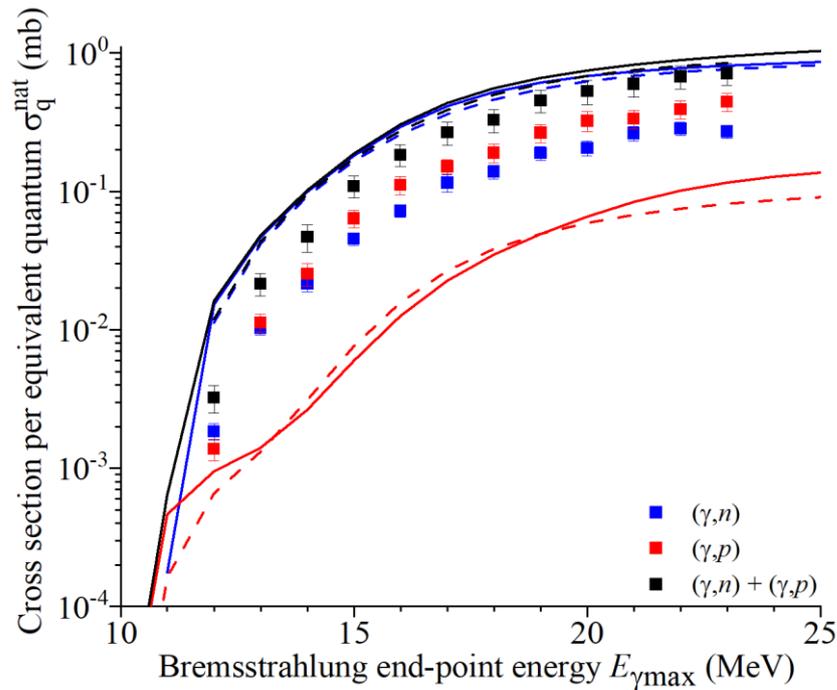

Figure 10. Cross section per equivalent quantum of $^{106}Cd(\gamma, n)^{105}Cd$ and $^{106}Cd(\gamma, p)^{105}Ag$ reactions as a function of bremsstrahlung end-point energy from the present work (solid rectangles) and simulated values using the CMPR (solid lines) and TALYS code (dashed lines)

2. $^{112}Cd(\gamma, p)^{111}Ag$ reaction

For the $^{112}Cd(\gamma, p)^{111}Ag$ reaction only one previous experimental data sets in 23 MeV [3]; this was found to be lower than the theoretical values obtained using the CMPR code as shown in Fig. 11. In Fig. 11, it is shown that the currently measured and theoretical values based on the CMPR code are in good agreement, in terms of not only shape but also magnitude. TALYS predicts results that are about almost 5 times lower than CMPR, and this difference increases rapidly with increasing energy.

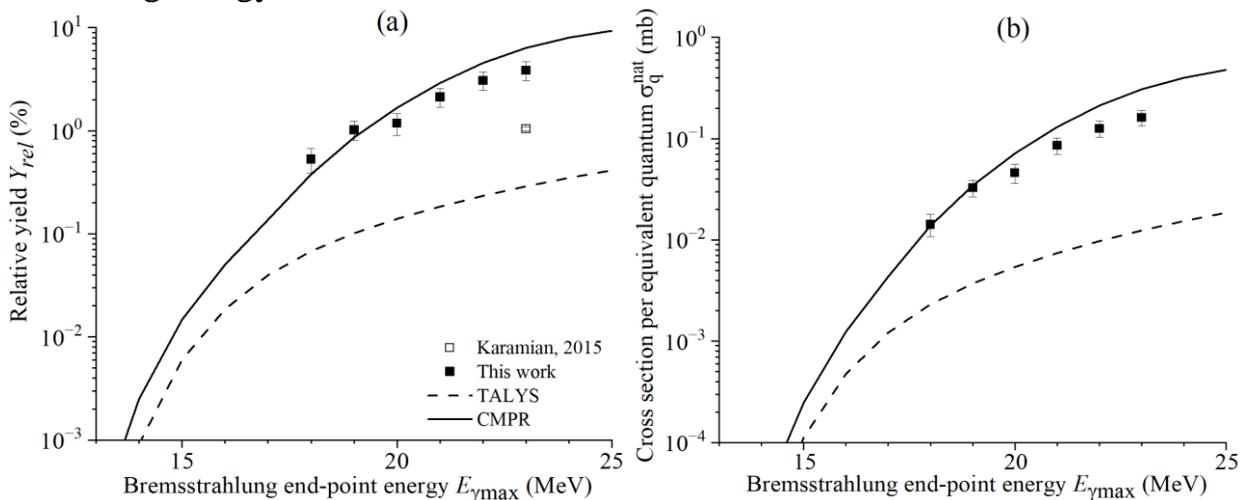

Figure 11. Relative yields (*a*) and cross section per equivalent quantum (*b*) of reaction $^{112}Cd(\gamma, p)^{111}Ag$ as a function of bremsstrahlung end-point energy from the present work (solid rectangles), literature data [3] (open rectangle) as well as simulated values using the CMPR (solid lines) and TALYS code (dashed lines)

### 3. $^{113}$Cd($\gamma$, $p$)$^{112}$Ag reaction

For the $^{113}$Cd($\gamma$, $p$)$^{112}$Ag reaction only one previous experimental data sets in 23 MeV [3]; this was found to be lower than the theoretical values obtained using the CMPR code as shown in Fig. 12. In Fig. 12, it is shown that the measured values for the reaction are higher than the calculated values; however, they are the closest to the values calculated using the CMPR code. TALYS predicts results that are about almost 5 times lower than CMPR, and this difference increases rapidly with increasing energy.

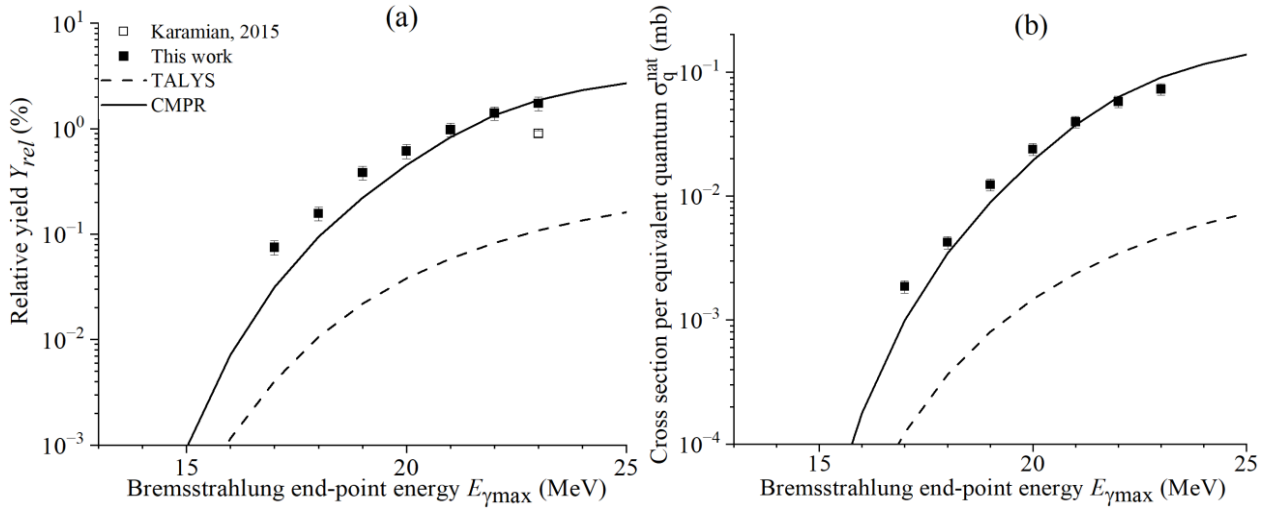

Figure 12. Relative yields (*a*) and cross section per equivalent quantum (*b*) of reaction $^{113}$Cd($\gamma$, $p$)$^{112}$Ag as a function of bremsstrahlung end-point energy from the present work (solid rectangles), literature data [3] (open rectangle) as well as simulated values using the CMPR (solid lines) and TALYS code (dashed lines)

### 4. $^{114}$Cd($\gamma$, $p$)$^{113}$Ag reaction

For the $^{114}$Cd($\gamma$, $p$)$^{113}$Ag reaction only one previous experimental data sets in 23 MeV [3]; this was found to be lower than the theoretical values obtained using the CMPR code as shown in Fig. 13. In Fig. 13, it is shown that the measured values for the reaction are higher than the calculated values; however, they are the closest to the values calculated using the CMPR code. TALYS predicts results that are about almost 5 times lower than CMPR, and this difference increases rapidly with increasing energy.

As can be seen in Fig.11-13, it is clear that a discrepancy in experimental data on the reactions $^{nat}$Cd($\gamma$, $p$)$^{111-113}$Ag is observed, our results are higher than literature data [3]. The large difference between the results of the work [3] and the present ones on the reactions $^{nat}$Cd($\gamma$, $p$)$^{111-113}$Ag might arise from the difference in the measuring duration of the irradiated targets.

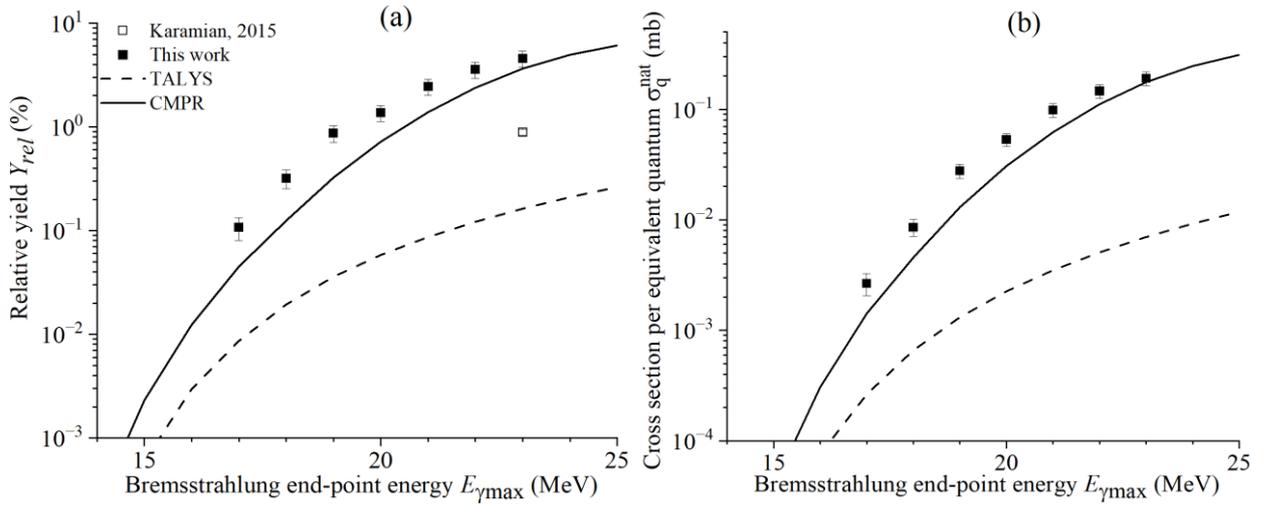

Figure 13. Relative yields (*a*) and cross section per equivalent quantum (*b*) of reaction $^{114}$Cd($\gamma$, *p*)$^{113}$Ag as a function of bremsstrahlung end-point energy from the present work (solid rectangles), literature data [3] (open rectangle) as well as simulated values using the CMPR (solid lines) and TALYS code (dashed lines)

5. $^{116}$Cd($\gamma$, *p*)$^{115}$Ag reaction

The measurements from the reaction were only compared with the theoretical values due to unavailability of published data. In Fig. 14, it is shown that the measured values for the reaction are higher than the calculated values; however, they are the closest to the values calculated using the CMPR code. TALYS predicts results approximately 16 times lower than CMPR.

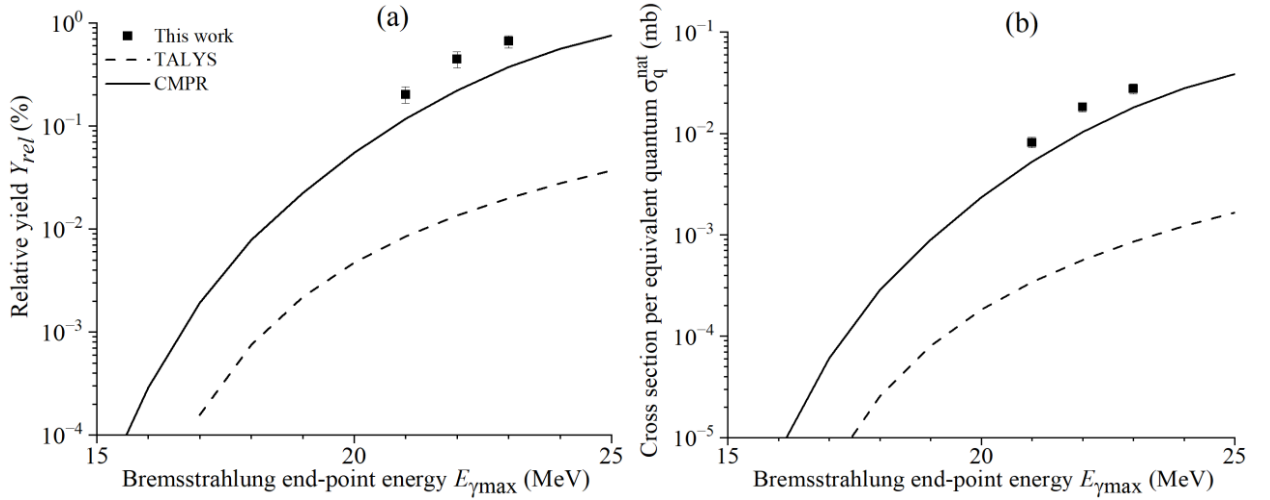

Figure 14. Relative yields (*a*) and cross section per equivalent quantum (*b*) of reaction $^{116}$Cd($\gamma$, *p*)$^{115}$Ag as a function of bremsstrahlung end-point energy from the present work (solid rectangles) and simulated values using the CMPR (solid lines) and TALYS code (dashed lines)

### B. Photonuclear reactions on tellurium isotopes
### B1. Photoneutron reactions

Nine tellurium isotopes are directly generated by $^{nat}$Te($\gamma$, *n*) reactions when natural tellurium is irradiated with bremsstrahlung radiation with end-point energy of 10-23 MeV. In this study, the relative yields and cross sections per equivalent quantum of the $^{nat}Te(\gamma, n)^{119g,119m,121g,121m,123m,125m,127,129g,129m}Te$ reactions at the bremsstrahlung end-point energies of 10-23 MeV are determined and presented in Fig. 15-20.

1. $^{120}Te(\gamma, n)^{119}Te$ reaction

The measured results for the $^{120}Te(\gamma, n)^{119m,g}Te$ reactions are compared with the theoretical values obtained with the TALYS and CMPR codes, as shown in Fig. 15. There is only one literature data for the $^{120}Te(\gamma, n)^{119}Te$ reaction [22] in the energy range of γ quantum 8.03-26.46 MeV. The theoretical values from both the TALYS and CMPR codes align with the literature data; yet, our results are inferior to them, as shown in Fig. 15.

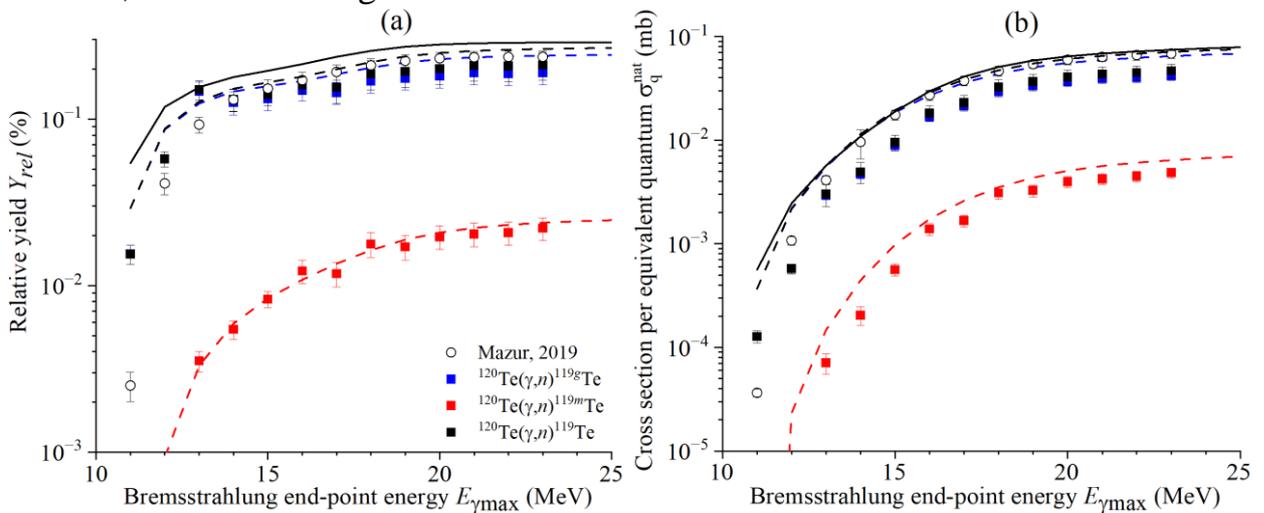

Figure 15. Relative yields (*a*) and cross section per equivalent quantum (*b*) of reaction $^{120}Te(\gamma, n)^{119}Te$ as a function of bremsstrahlung end-point energy from the present work (solid rectangles), literature data [22] (open circles) as well as simulated values using the CMPR (solid lines) and TALYS code (dashed lines)

2. $^{122}Te(\gamma,n)^{121}Te$ and $^{123}Te(\gamma,2n)^{121}Te$ reactions

The measured results for the $^{122}Te(\gamma,n)^{121}Te$ and $^{123}Te(\gamma,2n)^{121}Te$ reactions are compared with the theoretical values obtained with the TALYS and CMPR codes, as shown in Fig. 16. The measured values for the reaction are lower than the calculated values, as seen in Fig. 16.

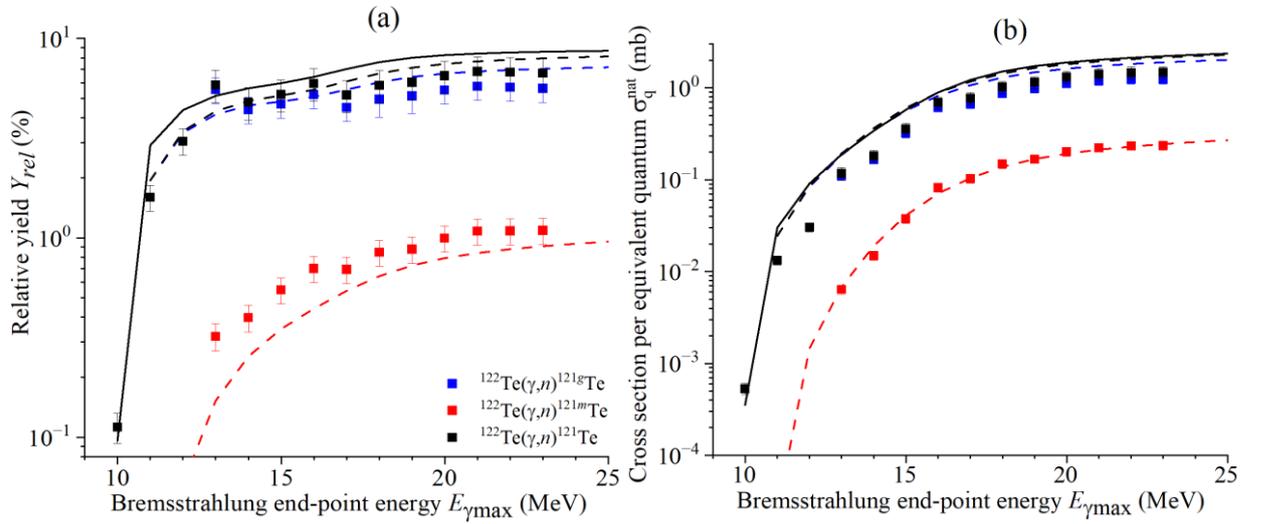

Figure 16. Relative yields (*a*) and cross section per equivalent quantum (*b*) of the $^{122}$Te($\gamma$,n)$^{121}$Te and $^{123}$Te($\gamma$,2n)$^{121}$Te reactions as a function of bremsstrahlung end-point energy from the present work (solid rectangles) as well as simulated values using the CMPR (solid lines) and TALYS code (dashed lines)

3. $^{123}$Te($\gamma$, $\gamma$`)$^{123m}$Te, $^{124}$Te($\gamma$, n)$^{123m}$Te and $^{125}$Te($\gamma$, 2n)$^{123m}$Te reactions

The measured results for the $^{123}$Te($\gamma$, $\gamma$`)$^{123m}$Te, $^{124}$Te($\gamma$, n)$^{123m}$Te and $^{125}$Te($\gamma$, 2n)$^{123m}$Te reactions are compared with the theoretical values obtained with the TALYS code, as shown in Fig. 17. The measured values for the reaction are lower than the calculated values, as seen in Fig. 17.

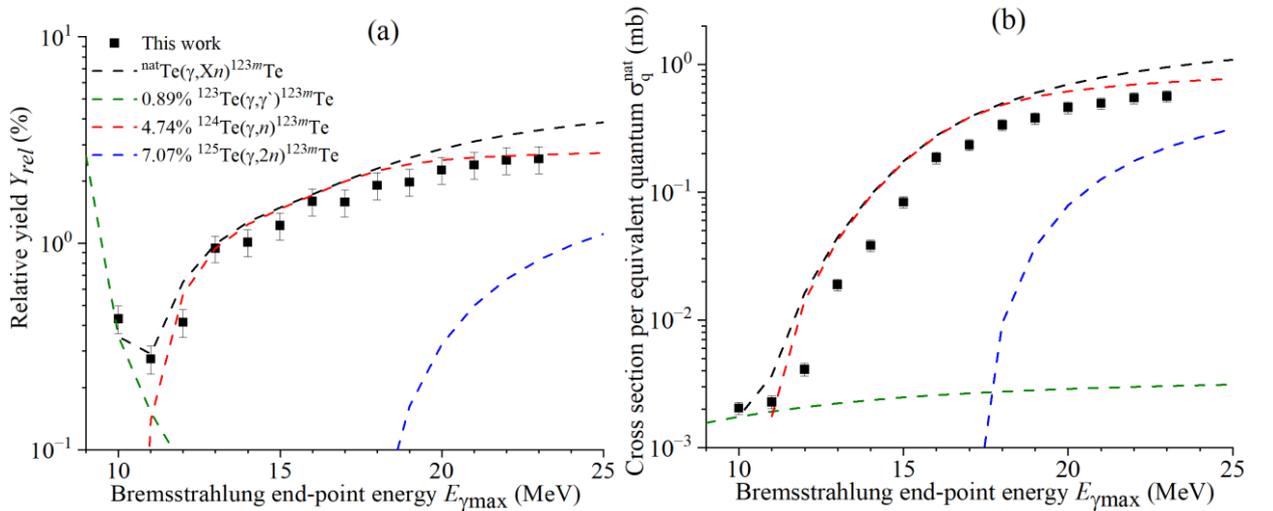

Figure 17. Relative yields (*a*) and cross section per equivalent quantum (*b*) of $^{123}$Te($\gamma$, $\gamma$`)$^{123m}$Te, $^{124}$Te($\gamma$, n)$^{123m}$Te and $^{125}$Te($\gamma$, 2n)$^{123m}$Te reactions as a function of bremsstrahlung end-point energy from the present work (solid rectangles) and simulated values using TALYS code

4. $^{125}$Te($\gamma$, $\gamma$`)$^{125m}$Te and $^{126}$Te($\gamma$, n)$^{125m}$Te reactions

The measured results for the $^{125}$Te($\gamma$, $\gamma$`)$^{125m}$Te and $^{126}$Te($\gamma$, n)$^{125m}$Te reactions are compared with the theoretical values obtained with the TALYS code, as shown in Fig. 18. As can be seen in Fig. 18a, the measured values for the reaction are higher than the calculated values. This may be due to the fact that the nucleus $^{125m}$Te emits a $\gamma$-ray with low intensity (109.28 keV (0.28%)). However, with exception data in

10-12 MeV, there is a good agreement between measured data and TALYS value in Fig. 18b.

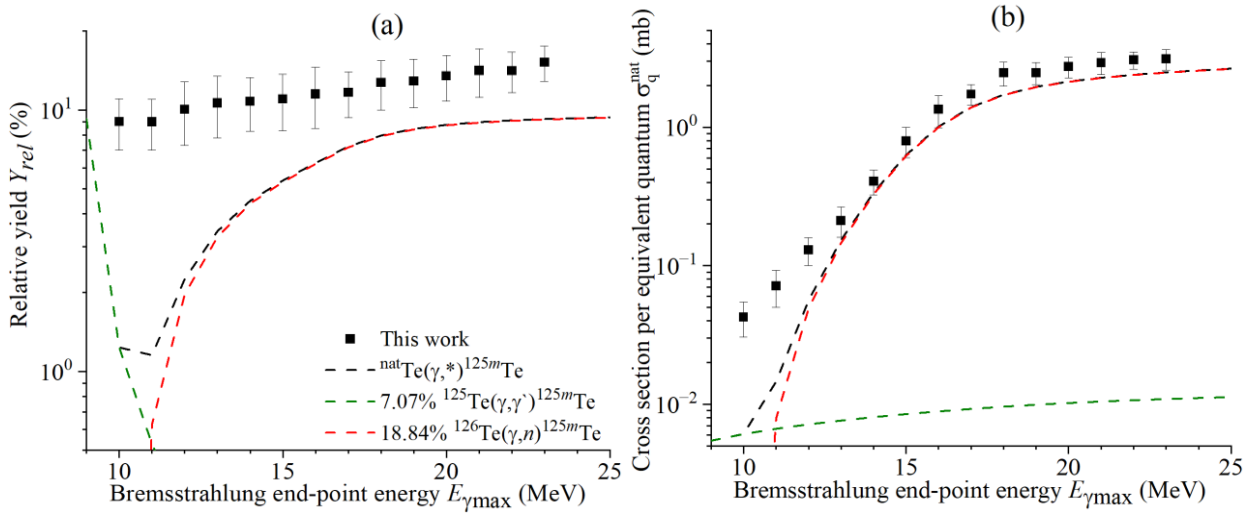

Figure 18. Relative yields (*a*) and cross section per equivalent quantum (*b*) of $^{125}$Te($\gamma$, $\gamma$`)$^{125m}$Te and $^{126}$Te($\gamma$, $n$)$^{125m}$Te reactions as a function of bremsstrahlung end-point energy from the present work (solid rectangles) and simulated values using TALYS code

5. $^{128}$Te($\gamma$,$n$)$^{127}$Te reaction

The measured results for the $^{128}$Te($\gamma$,$n$)$^{127}$Te reaction are compared with the theoretical values obtained with the TALYS, as shown in Fig. 19. The measured values for the reaction are lower than the calculated values, as seen in Fig. 19. There is a good agreement between literature data and theoretical calculations.

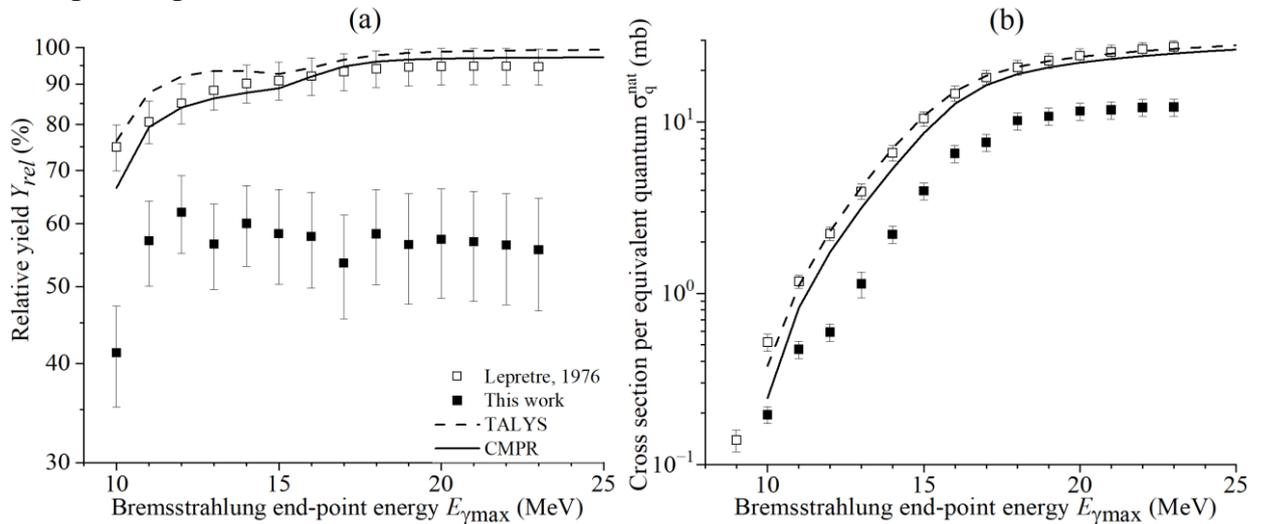

Figure 19. Relative yields (*a*) and cross section per equivalent quantum (*b*) of $^{128}$Te($\gamma$, $n$)$^{127}$Te reaction as a function of bremsstrahlung end-point energy from the present work (solid rectangles), literature data [22] (open rectangles) as well as simulated values using the CMPR (solid lines) and TALYS code (dashed lines)

6. $^{130}$Te($\gamma$, $n$)$^{129}$Te reaction

The measured results for the $^{120}$Te($\gamma$, $n$)$^{119m,g}$Te reactions are compared with the theoretical values obtained with the TALYS and CMPR codes, as shown in Fig.

20. There is only one literature data for the $^{130}$Te(γ, n)$^{129}$Te reaction [22] in the energy range of γ quantum 8.03-26.46 MeV. It is clear that the theoretical values from both the TALYS and CMPR codes are in agreement with the literature data, but our results are below them, as shown in Fig. 20.

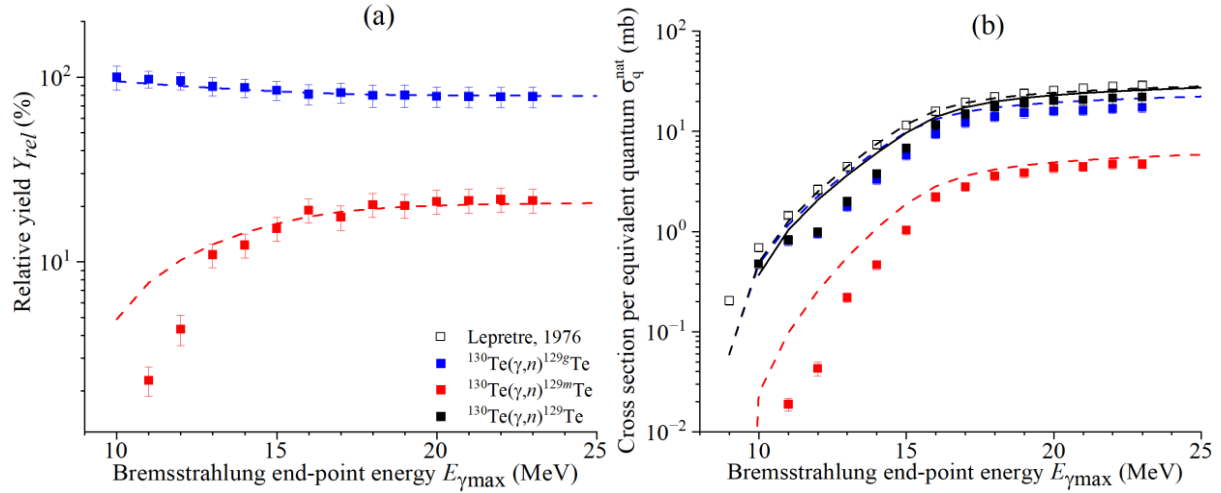

Figure 20. Relative yields (*a*) and cross section per equivalent quantum (*b*) of reaction $^{130}$Te(γ, n)$^{129}$Te as a function of bremsstrahlung end-point energy from the present work (solid rectangles), literature data [22] (open rectangles) as well as simulated values using the CMPR (solid line) and TALYS code (dashed lines)

### B2. Photoproton reactions

Four stibium ($^{122}$Sb, $^{124}$Sb, $^{127}$Sb, $^{129}$Sb) radioisotopes were directly produced by $^{nat}$Te(γ, p) reactions. In this study, the relative yields and cross section per equivalent quantum of the $^{nat}$Te(γ, p)$^{122,124,127,129}$Sb reactions at the bremsstrahlung end-point energies of 10-23 MeV are determined for the first time and presented in Fig. 21-25.

1. $^{123}$Te(γ, p)$^{122}$Sb and $^{124}$Te(γ, np)$^{122}$Sb reactions

Fig. 21 displays the measured data as well as the computed values. It is evident from the Fig. 21 that there is good agreement between the theoretical values on the basis of CMPR only for 22 and 23 MeV. The remaining experimental points are almost 100 times larger than theoretical calculations.

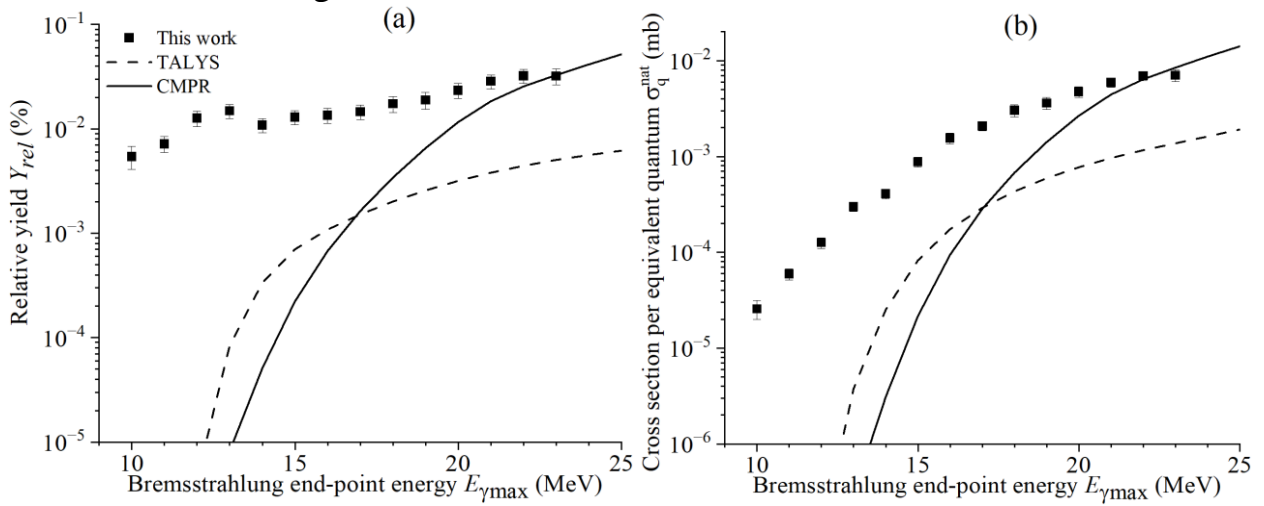

Figure 21. Relative yields (*a*) and cross section per equivalent quantum (*b*) of the $^{123}$Te($\gamma$, $p$)$^{122}$Sb and $^{124}$Te($\gamma$, $np$)$^{122}$Sb reactions as a function of bremsstrahlung end-point energy from the present work (solid rectangles) as well as simulated values using the CMPR (solid lines) and TALYS code (dashed lines)

Since the threshold of the reaction $^{124}$Te($\gamma$, $d$)$^{122}$Sb is 15.33 MeV, then we can assume that up to 19 MeV (taking into account the Coulomb barrier) the nucleus $^{122}$Sb is formed as a result of the reaction $^{123}$Te($\gamma$, $p$). Fig. 22 shows the ratios of the cross section per equivalent quantum $\sigma_{qexp}^{nat}/\sigma_{qtheory}^{nat}$ for the ($\gamma$, $p$) reaction on $^{123}$Te. As can be seen in Fig. 22, both models cannot describe the experimental points.

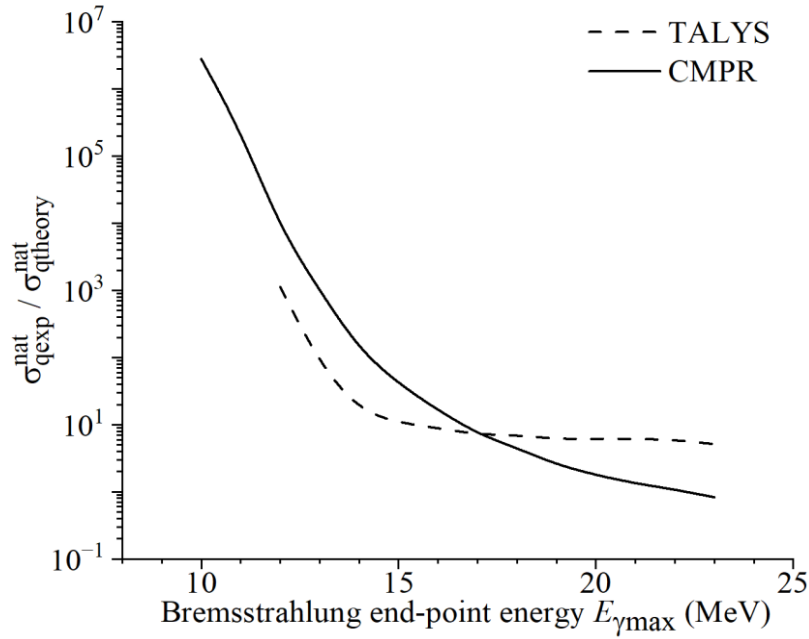

Figure 22. The ratio of the cross section per equivalent quantum $\sigma_{qexp}^{nat}/\sigma_{qtheory}^{nat}$ for the $^{123}$Te($\gamma$, $p$)$^{122}$Sb reaction

2. $^{125}$Te($\gamma$, $p$)$^{124}$Sb and $^{126}$Te($\gamma$, $np$)$^{124}$Sb reactions

The measured results for the $^{125}$Te($\gamma$, $p$)$^{124}$Sb and $^{126}$Te($\gamma$, $np$)$^{124}$Sb reactions are compared with the theoretical values obtained with the TALYS and CMPR codes, as shown in Fig. 23. It is clear that there is a discrepancy between theoretical calculations, and the experimental points are higher than the TALYS curve, but lower than the CMPR curve with an exception 19 MeV (this point is in good agreement with the TALYS calculation).

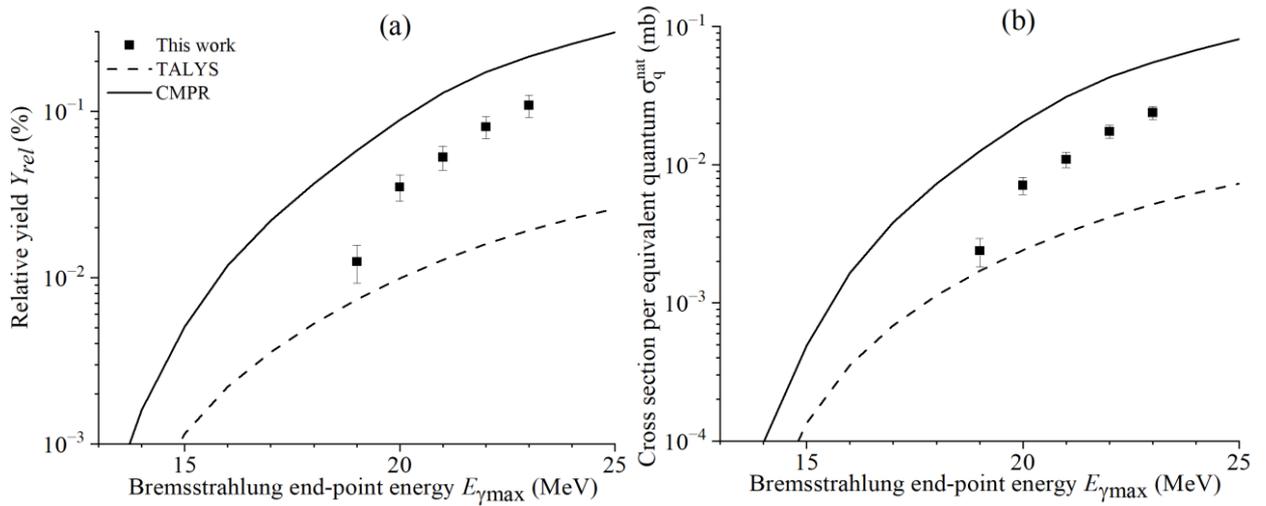

Figure 23. Relative yields (*a*) and cross section per equivalent quantum (*b*) of the $^{125}$Te($\gamma, p$)$^{124}$Sb and $^{126}$Te($\gamma, np$)$^{124}$Sb reactions as a function of bremsstrahlung end-point energy from the present work (solid rectangles) as well as simulated values using the CMPR (solid lines) and TALYS code (dashed lines)

3. $^{128}$Te($\gamma, p$)$^{127}$Sb reaction

The measured results for the $^{128}$Te($\gamma,p$)$^{127}$Sb reaction are compared with the theoretical values obtained with the TALYS and CMPR codes, as shown in Fig. 24. It is evident from the Fig. 24 that there is good agreement between the theoretical values on the basis of CMPR and currently measured values in terms of both form and magnitude.

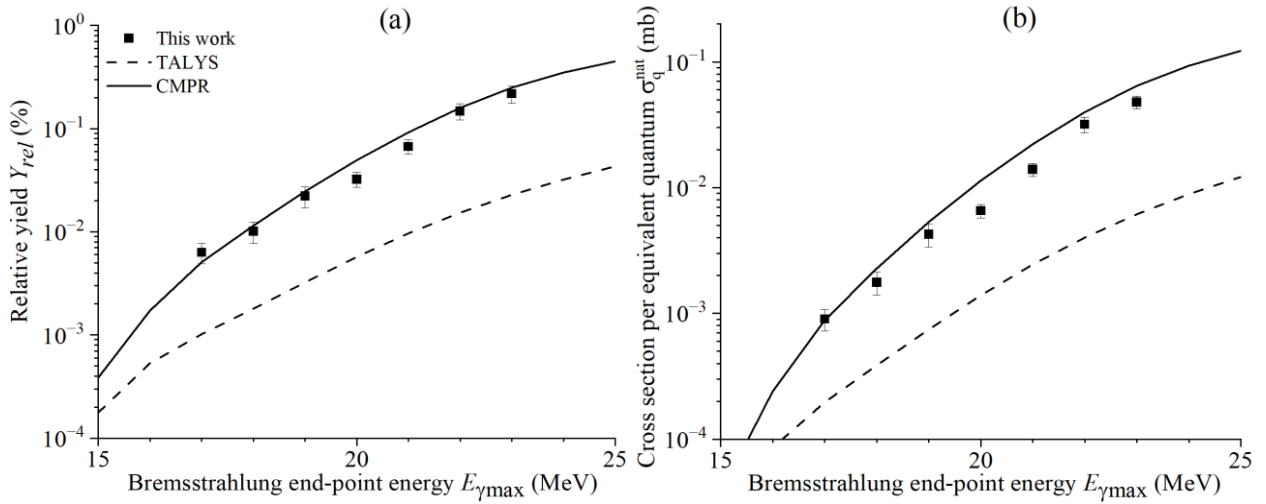

Figure 24. Relative yields (*a*) and cross section per equivalent quantum (*b*) of the reaction $^{128}$Te($\gamma, p$)$^{127}$Sb as a function of bremsstrahlung end-point energy from the present work (solid rectangles) as well as simulated values using the CMPR (solid lines) and TALYS code (dashed lines)

4. $^{130}$Te($\gamma, p$)$^{129}$Sb reaction

Fig. 25 displays the measured data as well as the computed values. It is clear that there is a discrepancy between theoretical calculations, and the experimental points are higher than the TALYS curve, but lower than the CMPR curve. The experimentally obtained results lie closer to the theoretical curve according to the CMPR.

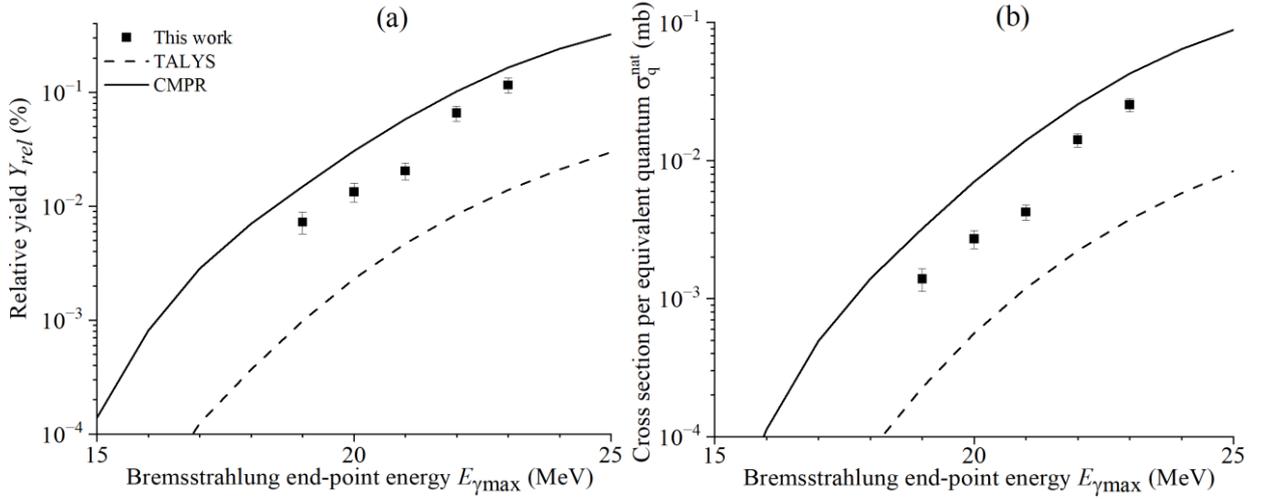

Figure 25. Relative yields (*a*) and cross section per equivalent quantum (*b*) of the reaction $^{130}$Te($\gamma$, $p$)$^{129}$Sb as a function of bremsstrahlung end-point energy from the present work (solid rectangles) as well as simulated values using the CMPR (solid lines) and TALYS code (dashed lines)

### C. Analysis of the results on ($\gamma$,$p$) reactions

Since we are irradiating natural mixtures of Cd and Te, this will give us the opportunity to study photonuclear reactions over a wide mass range with A=106-130. The cross section per equivalent quantum of the ($\gamma$,$p$) reaction in a case of monoisotopes $\sigma_q$ calculated on the basis of the CMPR and TALYS by equation (5) are contrasted against experimental data in Fig. 26. The data in Fig. 26 shows that the measured yields of the ($\gamma$,$p$) reactions on the isotopes $^{112,113,114,116}$Cd and $^{128,130}$Te agree within 30% with the results of the calculations based on the CMPR. The ($\gamma$,$p$) yields calculated on the basis of the TALYS code are underestimated with respect to experimental data by one order of magnitude. The reason behind this difference is that the TALYS code disregards special features of the decay of the $T_>$ GDR component, whose decay through the proton channel to low-lying states of the final nucleus is forbidden by isospin-selection rules. As a result, $T_>$ states decay through the proton channel with a higher probability. The cross section per equivalent quantum calculations for the reactions $^{114}$Cd($\gamma$,$p$)$^{113}$Ag and $^{128}$Te($\gamma$,$p$)$^{127}$Sb using TALYS parameters on the basis of the models of the nuclear level densities and $\gamma$-ray strength functions and accounting of isospin splitting in CMPR are given in Appendix 2.

The measured ($\gamma$,$p$) yields are a few percent of the ($\gamma$,$n$) yields for all nuclei, with the exception of the $^{106}$Cd nucleus, for which the yield of the reaction $^{106}$Cd($\gamma$,$p$)$^{105}$Cd is commensurate with the yield of the reaction $^{106}$Cd($\gamma$,$n$)$^{105}$Cd, but this contradicts the results of the theoretical calculations in [24,25]. In the case of the $^{123}$Te($\gamma$,$p$)$^{122}$Sb reaction, the situation is completely unclear. In all experiments, the spectra show a line at 564 keV. However, theoretical calculations on the basis of TALYS and CMPR are several orders of magnitude smaller.

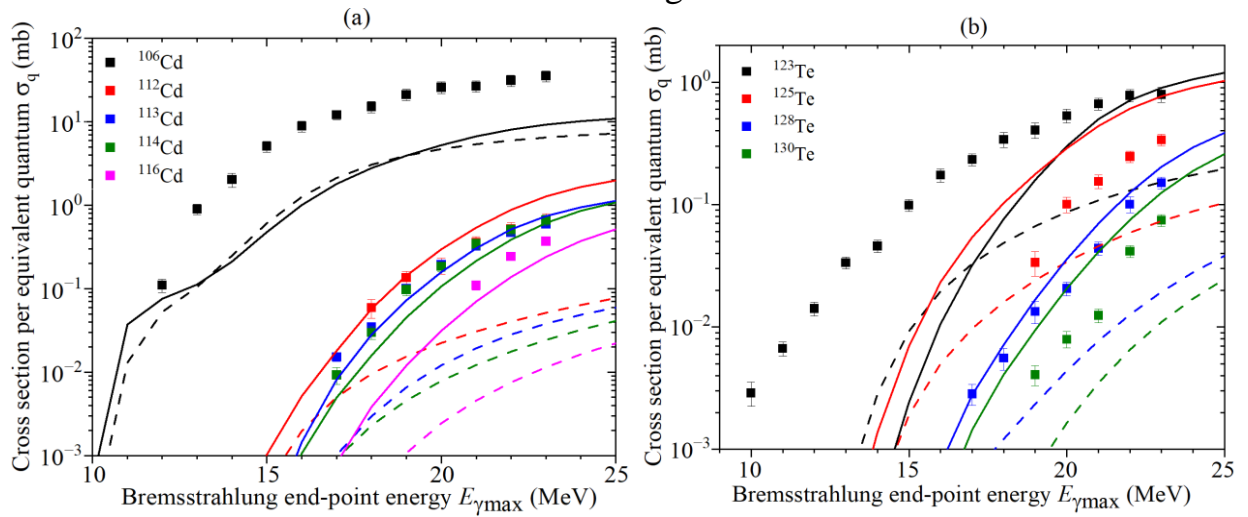

Figure 26. Cross section per equivalent quantum of the (γ,p) reaction in a case of monoisotopes $\sigma_q$ for the stable isotopes of cadmium (a) and tellurium (b)

For a more thorough analysis of the obtained results, the ratios of the reaction yield (γ,p) to the reaction yield (γ,n) were calculated for the nuclei $^{106}$Cd ($^{105}$Ag/$^{105}$Cd), $^{116}$Cd ($^{115}$Ag/$^{115}$Cd), $^{128}$Te ($^{127}$Sb/$^{127}$Te) and $^{130}$Te ($^{129}$Sb/$^{129}$Te). Also the ratios of yields (γ,p)/(γ,n) are calculated for the nucleus $^{74}$Se on the basis of the experimental results of our previously work [41]. Figure 27a shows the ratio of the reaction yield (γ,p) to the reaction yield (γ,n) depending on the electron energy of the accelerator. As shown in Fig. 27a the ratios of yields (γ,p)/(γ,n) are almost equal to 1 for bypassed nuclei $^{74}$Se and $^{106}$Cd. It can be seen that with the exception of the points for the nucleus $^{106}$Cd, the experimental points are consistent with the calculated curves of CMPR. In the case of $^{106}$Cd ($^{105}$Ag/$^{105}$Cd), the CMPR and TALYS cannot describe the experimental results. The reason for the observed discrepancy between theory and experiment may be explained by the fact that statistical models of photonuclear reactions do not take into account the individual structural features of Cd isotopes.

Figure 27b shows the ratios of yields (γ,p)/(γ,n) as a function of the proton-neutron ratio $N/Z$ at the electron energy of the accelerator 23 MeV. The literature data were calculated using experimentally measured reaction cross sections (γ,p) and (γ,n) on the nuclei $^{90}$Zr [42,43], $^{89}$Y [43,44], $^{103}$Rh [1,45], $^{112}$Sn [46,47] and $^{160}$Gd [48]. With the exception of the points at $N/Z = 1.176$ ($^{74}$Se) and 1.208 ($^{106}$Cd), the experimental points are consistent with the calculated curve on the basis of CMPR. The ratio of yields (γ,p)/(γ,n) decreases with increasing proton-neutron ratio $N/Z$.

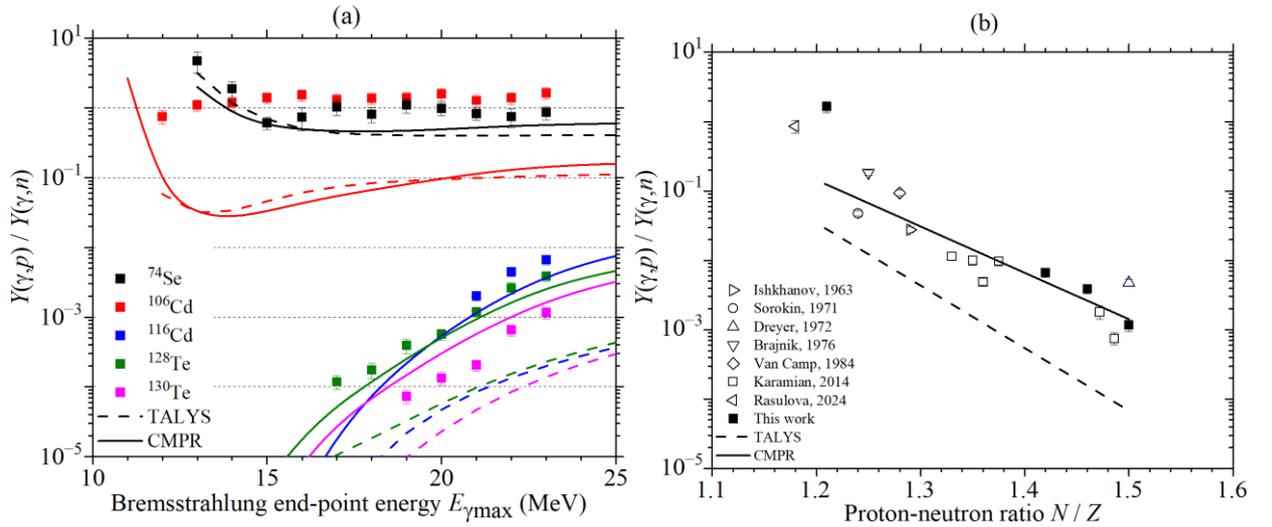

Fig. 27. The ratio of the reactions' yields $(\gamma,p)/(\gamma,n)$ as a function of:
(a) the electron energy of the accelerator and (b) the proton-neutron ratio $N/Z$ at the electron energy of the accelerator 23 MeV

### D. Isomeric ratios for the pairs $^{116}$Cd$(\gamma,n)^{115m,g}$Cd, $^{120}$Te$(\gamma,n)^{119m,g}$Te, $^{122}$Te$(\gamma,n)^{121m,g}$Te and $^{130}$Te$(\gamma,n)^{129m,g}$Te

Based on the measured experimental yields of the metastable and ground states from Table 1-2 in Appendix 1, we obtained the isomeric yield ratio ($IR=\sigma_h/\sigma_l$) of $^{115g}$Cd (nuclear spin=1/2$^+$) and $^{115m}$Cd (nuclear spin=11/2$^-$) in the $^{nat}$Cd$(\gamma,n)$ reactions, which are given in Table 3 for various bremsstrahlung end-point energies. The photon induced isomeric ratio values from this study, the literature data in the GDR region [7-15] are shown in Fig. 28a. As seen in Fig. 28a, the experimental isomeric ratio values in the $^{116}$Cd$(\gamma, n)$ reaction are in agreement with the theoretical values. Furthermore, Fig. 28a shows that the isomeric ratio values of $^{115m,g}$Cd increase with increasing excitation energy.

In the experiments we have registered three isomeric pairs of tellurium $^{119,121,129}$Te. All of the isomeric states are with nuclear spin 11/2$^-$. Based on the measured experimental yields of the metastable and ground states, we obtained the isomeric yield ratio ($IR=\sigma_h/\sigma_l$) for all of them, which are given in Fig. 28b,c,d for various bremsstrahlung end-point energies.

For the reaction $^{120}$Te$(\gamma,n)^{119}$Te the photon induced isomeric ratio values from this study, the literature data in the GDR region [10,19,20] are shown in Fig. 27b. As seen in Fig. 28b, our experimental isomeric ratio values in the $^{120}$Te$(\gamma, n)$ reaction are in agreement with the TALYS curve. Starting from 15 MeV, there is a scatter in our and literature data in isomeric ratios. Moreover, the figure shows that the isomeric ratio values of $^{119m,g}$Te increase with increasing excitation energy.

For the reaction $^{122}$Te$(\gamma,n)^{121}$Te the photon induced isomeric ratio values from this study, the literature data in the GDR region [10,11,19,20] are shown in Fig. 28c. As seen in Fig. 28c, our experimental values of isomeric ratios in the reaction $^{nat}$Te$(\gamma,n)^{121m,g}$Te are consistent with the literature data, but they diverge from the TALYS curve starting from 11 MeV. Additionally, As the excitation energy increases, the isomeric ratio values of $^{121m,g}$Te rise, as the figure illustrates.

For the reaction $^{130}$Te($\gamma,n$)$^{129}$Te the photon induced isomeric ratio values from this study, the literature data in the GDR region [10-12,17,19,21] are shown in Fig. 28d. As seen in Fig. 28d, our experimental isomeric ratio values in the $^{130}$Te($\gamma,n$) reaction are in agreement with the TALYS curve. Starting from 14 MeV, there is a scatter in our and literature data in isomeric ratios. Furthermore, the figure shows that the isomeric ratio values of $^{129m,g}$Te increase with increasing excitation energy.

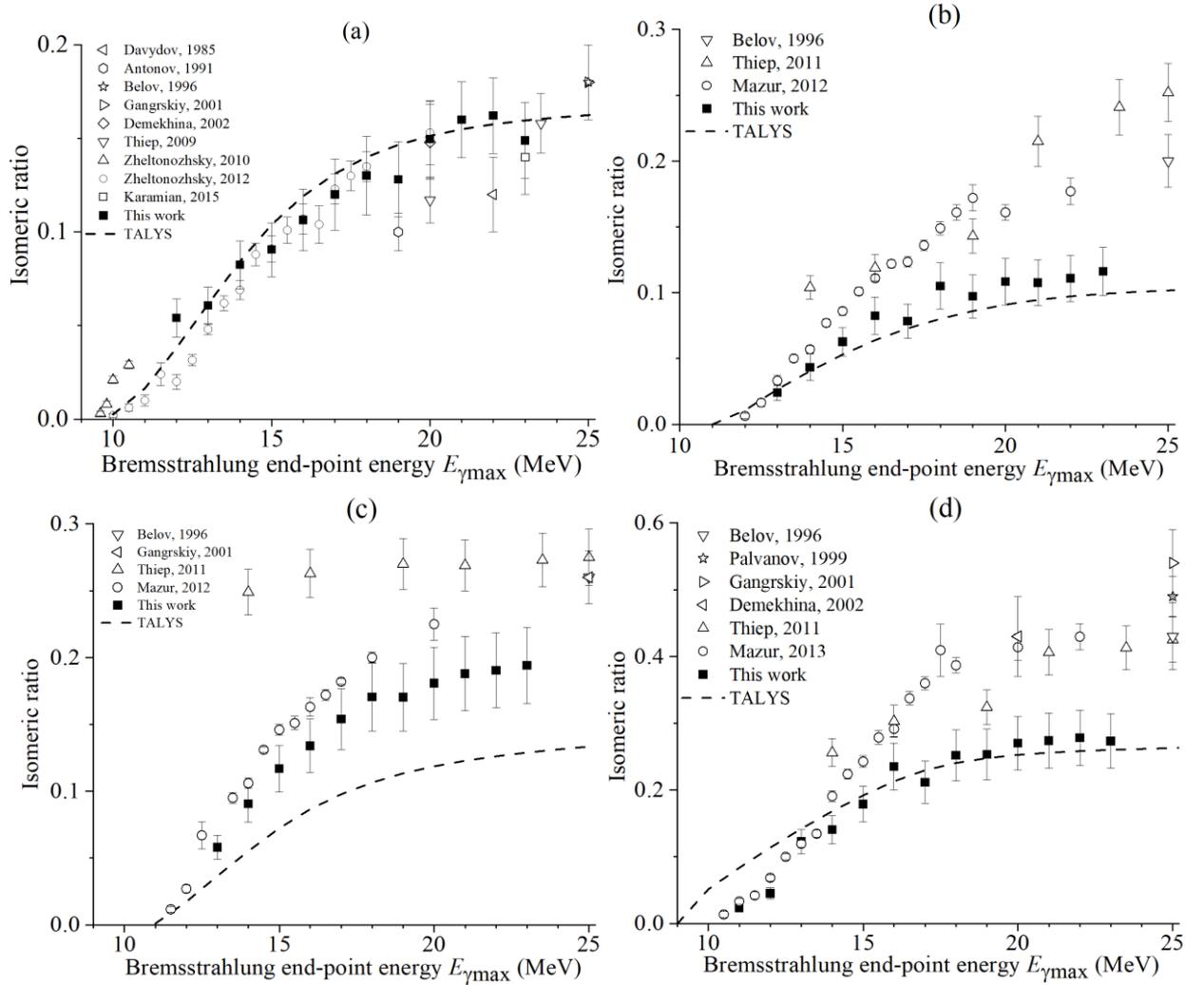

Figure 28. Isomeric yield ratios of the pairs $^{116}$Cd$(\gamma,n)^{115m,g}$Cd (a), $^{120}$Te$(\gamma,n)^{119m,g}$Te (b), $^{122}$Te$(\gamma,n)^{121m,g}$Te (c) and $^{130}$Te$(\gamma,n)^{129m,g}$Te (d) as a function of the bremsstrahlung end-point energy from the present work (solid squares), literature data and simulated values using TALYS code (dashed lines)

Table 3. Isomeric ratios of the $^{116}$Cd$(\gamma,n)^{115}$Cd and $^{nat}$Te$(\gamma,n)^{119,121,129}$Te reactions at the bremsstrahlung end-point energies of 11–23 MeV from the present work

| Energy (MeV) | $^{116}$Cd$(\gamma,n)^{115}$Cd | $^{120}$Te$(\gamma,n)^{119}$Te | $^{122}$Te$(\gamma,n)^{121}$Te | $^{130}$Te$(\gamma,n)^{129}$Te |
|---|---|---|---|---|
| 11 | | | | 0.023 ± 0.004 |
| 12 | 0.05 ± 0.01 | | | 0.045 ± 0.008 |
| 13 | 0.06 ± 0.01 | 0.024 ± 0.006 | 0.06 ± 0.01 | 0.12 ± 0.02 |
| 14 | 0.08 ± 0.01 | 0.04 ± 0.01 | 0.09 ± 0.01 | 0.14 ± 0.02 |
| 15 | 0.09 ± 0.01 | 0.06 ± 0.01 | 0.12 ± 0.02 | 0.18 ± 0.03 |
| 16 | 0.11 ± 0.02 | 0.08 ± 0.01 | 0.13 ± 0.02 | 0.23 ± 0.03 |
| 17 | 0.12 ± 0.02 | 0.08 ± 0.01 | 0.15 ± 0.02 | 0.21 ± 0.03 |
| 18 | 0.13 ± 0.02 | 0.11 ± 0.02 | 0.17 ± 0.03 | 0.25 ± 0.04 |
| 19 | 0.13 ± 0.02 | 0.10 ± 0.02 | 0.17 ± 0.03 | 0.25 ± 0.04 |
| 20 | 0.15 ± 0.02 | 0.11 ± 0.02 | 0.18 ± 0.03 | 0.27 ± 0.04 |
| 21 | 0.16 ± 0.02 | 0.11 ± 0.02 | 0.19 ± 0.03 | 0.27 ± 0.04 |
| 22 | 0.16 ± 0.02 | 0.11 ± 0.02 | 0.19 ± 0.03 | 0.28 ± 0.04 |
| 23 | 0.15 ± 0.02 | 0.12 ± 0.02 | 0.19 ± 0.03 | 0.27 ± 0.04 |

## 5. Conclusion

The present study addressed the measurements of relative yields and cross section per equivalent quantum for the photonuclear reactions on a natural mixture of cadmium and tellurium using bremsstrahlung end-point energies of 10 to 23 MeV. The bremsstrahlung photon flux was computed in the Geant4.11.1 code. The experimental results were compared with calculations using the TALYS model with the standard parameters and the CMPR. For the photoneutron reactions on the nuclei $^{112,116}$Cd($\gamma$,n) and $^{120,122,124,126,130}$Te, a good agreement was observed between the experimental data and theoretical calculations.

The measured ($\gamma$,p) yields are a few percent of the ($\gamma$,n) yields for all nuclei, with the exception of the bypassed nucleus $^{106}$Cd, for which the yield of the reaction $^{106}$Cd($\gamma$,p)$^{105}$Cd is commensurate with the yield of the reaction $^{106}$Cd($\gamma$,n)$^{105}$Cd, but this contradicts the results of the theoretical calculations. For the reaction $^{123}$Te($\gamma$,p)$^{122}$Sb experimental points are almost 100 times larger than theoretical calculations.

On the heavy isotopes $^{112,113,114,116}$Cd and $^{128,130}$Te, the experimental results agree with theoretical relative yields calculated using the CMPR. Including isospin splitting in the CMPR allows for much better description of experimental data on reactions with proton escape in the energy range from 12 to 23 MeV. Therefore, taking into account isospin splitting is necessary for a more reasonable description of the GDR decay. The study of photonuclear reactions on cadmium and tellurium isotopes is important for understanding the formation and decay of bypassed nuclei during nucleosynthesis.

# APPENDIX 1: THE DATA THAT SUPPORT THE FINDINGS OF THIS ARTICLE

In this Appendix, the tabulated experimental results discussed in Sec. 4 are presented.

Table 1. Experimental results of $^{nat}Cd(\gamma,n)$ and $^{nat}Cd(\gamma,p)$ reactions

| Reaction | $E_{\gamma max}$ (MeV) | $Y_{rel}$ (%) | $\sigma_q^{nat}$ (mb) |
|---|---|---|---|
| $^{106}Cd(\gamma, n)^{105}Cd$ | 12 | 0.7 ± 0.1 | 0.0018 ± 0.0003 |
| | 13 | 1.9 ± 0.3 | 0.010 ± 0.001 |
| | 14 | 3.1 ± 0.5 | 0.022 ± 0.003 |
| | 15 | 3.3 ± 0.4 | 0.045 ± 0.005 |
| | 16 | 3.4 ± 0.5 | 0.072 ± 0.008 |
| | 17 | 4.6 ± 0.6 | 0.11 ± 0.02 |
| | 18 | 5.1 ± 0.7 | 0.14 ± 0.02 |
| | 19 | 5.9 ± 0.8 | 0.19 ± 0.02 |
| | 20 | 5.3 ± 0.8 | 0.21 ± 0.02 |
| | 21 | 6.5 ± 0.9 | 0.26 ± 0.03 |
| | 22 | 6.9 ± 0.9 | 0.28 ± 0.03 |
| | 23 | 6.7 ± 0.9 | 0.27 ± 0.03 |
| $^{108}Cd(\gamma, n)^{107}Cd$ | 12 | 1.9 ± 0.4 | 0.006 ± 0.001 |
| | 13 | 3.1 ± 0.5 | 0.017 ± 0.002 |
| | 14 | 5.0 ± 0.7 | 0.035 ± 0.004 |
| | 15 | 5.3 ± 0.6 | 0.073 ± 0.008 |
| | 16 | 6.1 ± 0.9 | 0.13 ± 0.02 |
| | 17 | 6.4 ± 0.9 | 0.16 ± 0.02 |
| | 18 | 7 ± 1 | 0.19 ± 0.02 |
| | 19 | 8 ± 1 | 0.26 ± 0.03 |
| | 20 | 7 ± 1 | 0.30 ± 0.03 |
| | 21 | 8 ± 1 | 0.32 ± 0.03 |
| | 22 | 9 ± 1 | 0.35 ± 0.04 |
| | 23 | 9 ± 1 | 0.37 ± 0.04 |
| $^{110}Cd(\gamma, n)^{109}Cd$ | 12 | 44 ± 7 | 0.13 ± 0.02 |
| | 13 | 50 ± 9 | 0.27 ± 0.05 |
| | 14 | 88 ± 14 | 0.61 ± 0.09 |
| | 15 | 91 ± 11 | 1.2 ± 0.2 |
| | 16 | 89 ± 10 | 1.9 ± 0.2 |
| | 17 | 96 ± 11 | 2.4 ± 0.3 |
| | 18 | 105 ± 13 | 2.8 ± 0.4 |
| | 19 | 120 ± 19 | 3.9 ± 0.5 |
| | 20 | 135 ± 23 | 5.3 ± 0.7 |
| | 21 | 135 ± 22 | 5.4 ± 0.7 |
| | 22 | 141 ± 22 | 5.8 ± 0.7 |
| | 23 | 144 ± 22 | 6.0 ± 0.7 |
| $^{112}Cd(\gamma, n)^{111m}Cd$ | 10 | 14 ± 2 | 0.006 ± 0.001 |
| | 11 | 6.2 ± 0.9 | 0.007 ± 0.001 |
| | 12 | 6.0 ± 0.9 | 0.015 ± 0.002 |
| | 13 | 6 ± 1 | 0.034 ± 0.006 |
| | 14 | 12 ± 2 | 0.082 ± 0.009 |
| | 15 | 12 ± 2 | 0.16 ± 0.02 |
| | 16 | 8.7 ± 0.9 | 0.19 ± 0.02 |
| | 17 | 13 ± 2 | 0.32 ± 0.03 |
| | 18 | 20 ± 3 | 0.54 ± 0.06 |
| | 19 | 23 ± 3 | 0.75 ± 0.08 |
| | 20 | 22 ± 3 | 0.9 ± 0.1 |
| | 21 | 26 ± 4 | 1.1 ± 0.1 |
| | 22 | 27 ± 4 | 1.1 ± 0.1 |
| | 23 | 28 ± 4 | 1.2 ± 0.1 |
| $^{116}Cd(\gamma, n)^{115g}Cd$ | 10 | 100 ± 11 | 0.06 ± 0.01 |
| | 11 | 100 ± 11 | 0.12 ± 0.01 |
| | 12 | 95 ± 10 | 0.27 ± 0.03 |
| | 13 | 94 ± 10 | 0.50 ± 0.05 |
| | 14 | 92 ± 10 | 0.64 ± 0.07 |
| | 15 | 92 ± 10 | 1.3 ± 0.2 |
| | 16 | 90 ± 9 | 1.9 ± 0.2 |
| | 17 | 89 ± 9 | 2.2 ± 0.2 |
| | 18 | 88 ± 9 | 2.4 ± 0.2 |
| | 19 | 89 ± 9 | 2.9 ± 0.3 |
| | 20 | 87 ± 9 | 3.4 ± 0.3 |
| | 21 | 86 ± 9 | 3.5 ± 0.4 |
| | 22 | 86 ± 9 | 3.5 ± 0.4 |
| | 23 | 87 ± 9 | 3.6 ± 0.4 |
| $^{116}Cd(\gamma, n)^{115m}Cd$ | 12 | 5 ± 1 | 0.015 ± 0.002 |
| | 13 | 6 ± 1 | 0.031 ± 0.004 |
| | 14 | 8 ± 1 | 0.053 ± 0.006 |
| | 15 | 8 ± 1 | 0.11 ± 0.02 |
| | 16 | 10 ± 1 | 0.20 ± 0.02 |
| | 17 | 11 ± 1 | 0.26 ± 0.03 |
| | 18 | 12 ± 2 | 0.31 ± 0.04 |
| | 19 | 11 ± 2 | 0.36 ± 0.04 |
| | 20 | 13 ± 2 | 0.51 ± 0.06 |
| | 21 | 14 ± 2 | 0.58 ± 0.07 |
| | 22 | 14 ± 2 | 0.57 ± 0.07 |
| | 23 | 13 ± 2 | 0.54 ± 0.07 |
| $^{116}Cd(\gamma, n)^{115}Cd$ | 10 | | 0.06 ± 0.01 |
| | 11 | | 0.12 ± 0.01 |
| | 12 | | 0.28 ± 0.03 |
| | 13 | | 0.53 ± 0.05 |
| | 14 | | 0.69 ± 0.07 |
| | 15 | | 1.4 ± 0.2 |
| | 16 | | 2.1 ± 0.2 |
| | 17 | | 2.5 ± 0.3 |
| | 18 | | 2.7 ± 0.3 |
| | 19 | | 3.2 ± 0.3 |
| | 20 | | 3.9 ± 0.4 |
| | 21 | | 4.0 ± 0.4 |
| | 22 | | 4.1 ± 0.4 |
| | 23 | | 4.2 ± 0.4 |

| Reaction | $E_{\gamma max}$ (MeV) | $Y_{rel}$ (%) | $\sigma_q$ (mb) |
|---|---|---|---|
| $^{106}$Cd($\gamma, p$)$^{105}$Ag | 12 | 0.5 ± 0.1 | 0.0014 ± 0.0002 |
| | 13 | 2.1 ± 0.4 | 0.011 ± 0.002 |
| | 14 | 3.7 ± 0.8 | 0.025 ± 0.005 |
| | 15 | 4.6 ± 0.7 | 0.06 ± 0.001 |
| | 16 | 5.2 ± 0.9 | 0.11 ± 0.02 |
| | 17 | 6.1 ± 0.9 | 0.15 ± 0.02 |
| | 18 | 7.0 ± 1.3 | 0.19 ± 0.03 |
| | 19 | 8.0 ± 1.5 | 0.26 ± 0.04 |
| | 20 | 8.3 ± 1.6 | 0.32 ± 0.05 |
| | 21 | 8.3 ± 1.5 | 0.33 ± 0.05 |
| | 22 | 9.6 ± 1.5 | 0.39 ± 0.06 |
| | 23 | 11 ± 2 | 0.44 ± 0.06 |
| $^{112}$Cd($\gamma, p$)$^{111}$Ag | 18 | 0.5 ± 0.1 | 0.014 ± 0.004 |
| | 19 | 1.0 ± 0.2 | 0.033 ± 0.006 |
| | 20 | 1.2 ± 0.3 | 0.05 ± 0.01 |
| | 21 | 2.1 ± 0.4 | 0.08 ± 0.01 |
| | 22 | 3.1 ± 0.6 | 0.13 ± 0.02 |
| | 23 | 3.8 ± 0.8 | 0.16 ± 0.03 |
| $^{113}$Cd($\gamma, p$)$^{112}$Ag | 17 | 0.07 ± 0.01 | 0.0019 ± 0.0002 |
| | 18 | 0.16 ± 0.02 | 0.0042 ± 0.0004 |
| | 19 | 0.38 ± 0.06 | 0.012 ± 0.001 |
| | 20 | 0.61 ± 0.09 | 0.024 ± 0.003 |
| | 21 | 0.98 ± 0.15 | 0.039 ± 0.004 |
| | 22 | 1.4 ± 0.2 | 0.058 ± 0.006 |
| | 23 | 1.7 ± 0.2 | 0.073 ± 0.008 |
| $^{114}$Cd($\gamma, p$)$^{113}$Ag | 17 | 0.11 ± 0.03 | 0.003 ± 0.001 |
| | 18 | 0.32 ± 0.06 | 0.009 ± 0.002 |
| | 19 | 0.9 ± 0.1 | 0.028 ± 0.004 |
| | 20 | 1.4 ± 0.2 | 0.053 ± 0.007 |
| | 21 | 2.4 ± 0.4 | 0.098 ± 0.014 |
| | 22 | 3.6 ± 0.6 | 0.15 ± 0.02 |
| | 23 | 4.5 ± 0.8 | 0.19 ± 0.03 |
| $^{116}$Cd($\gamma, p$)$^{115}$Ag | 21 | 0.20 ± 0.03 | 0.008 ± 0.001 |
| | 22 | 0.44 ± 0.08 | 0.018 ± 0.002 |
| | 23 | 0.66 ± 0.09 | 0.028 ± 0.003 |

Table 2. Experimental results of $^{nat}$Te($\gamma,n$) and $^{nat}$Te($\gamma,p$) reactions

| Reaction | $E_{\gamma max}$ (MeV) | $Y_{rel}$ (%) | $\sigma_q$ (mb) |
|---|---|---|---|
| $^{120}$Te($\gamma, n$)$^{119g}$Te | 11 | 0.015 ± 0.002 | 0.00013 ± 0.0002 |
| | 12 | 0.058 ± 0.006 | 0.00058 ± 0.0006 |
| | 13 | 0.15 ± 0.02 | 0.0029 ± 0.0003 |
| | 14 | 0.13 ± 0.02 | 0.0047 ± 0.0005 |
| | 15 | 0.13 ± 0.02 | 0.009 ± 0.001 |
| | 16 | 0.15 ± 0.02 | 0.017 ± 0.002 |
| | 17 | 0.14 ± 0.02 | 0.021 ± 0.002 |
| | 18 | 0.17 ± 0.03 | 0.029 ± 0.003 |
| | 19 | 0.17 ± 0.03 | 0.034 ± 0.004 |
| | 20 | 0.18 ± 0.03 | 0.037 ± 0.004 |
| | 21 | 0.19 ± 0.03 | 0.039 ± 0.004 |
| | 22 | 0.19 ± 0.03 | 0.040 ± 0.004 |

| Reaction | | | |
|---|---|---|---|
| | 23 | 0.19 ± 0.03 | 0.042 ± 0.004 |
| $^{120}$Te$(\gamma, n)^{119m}$Te | 13 | 0.004 ± 0.001 | 0.00007 ± 0.00002 |
| | 14 | 0.005 ± 0.001 | 0.00020 ± 0.00004 |
| | 15 | 0.008 ± 0.001 | 0.0006 ± 0.0001 |
| | 16 | 0.012 ± 0.002 | 0.0014 ± 0.0002 |
| | 17 | 0.011 ± 0.002 | 0.0017 ± 0.0002 |
| | 18 | 0.018 ± 0.003 | 0.0031 ± 0.0004 |
| | 19 | 0.017 ± 0.003 | 0.0033 ± 0.0004 |
| | 20 | 0.019 ± 0.003 | 0.0040 ± 0.0005 |
| | 21 | 0.020 ± 0.003 | 0.0042 ± 0.0005 |
| | 22 | 0.021 ± 0.003 | 0.0045 ± 0.0005 |
| | 23 | 0.022 ± 0.003 | 0.0048 ± 0.0005 |
| $^{120}$Te$(\gamma, n)^{119}$Te | 11 | 0.015 ± 0.002 | 0.00013 ± 0.0002 |
| | 12 | 0.058 ± 0.006 | 0.00058 ± 0.0006 |
| | 13 | 0.15 ± 0.02 | 0.003 ± 0.001 |
| | 14 | 0.13 ± 0.02 | 0.005 ± 0.001 |
| | 15 | 0.14 ± 0.02 | 0.009 ± 0.002 |
| | 16 | 0.16 ± 0.02 | 0.018 ± 0.003 |
| | 17 | 0.15 ± 0.03 | 0.023 ± 0.004 |
| | 18 | 0.18 ± 0.03 | 0.033 ± 0.005 |
| | 19 | 0.19 ± 0.04 | 0.037 ± 0.006 |
| | 20 | 0.20 ± 0.04 | 0.041 ± 0.007 |
| | 21 | 0.21 ± 0.04 | 0.043 ± 0.007 |
| | 22 | 0.21 ± 0.04 | 0.045 ± 0.007 |
| | 23 | 0.21 ± 0.04 | 0.046 ± 0.007 |
| $^{122}$Te$(\gamma, n)^{121g}$Te | 10 | 0.11 ± 0.02 | 0.0005 ± 0.0001 |
| | 11 | 1.6 ± 0.2 | 0.013 ± 0.001 |
| | 12 | 3.0 ± 0.5 | 0.030 ± 0.003 |
| | 13 | 5.5 ± 0.8 | 0.11 ± 0.01 |
| | 14 | 4.4 ± 0.7 | 0.17 ± 0.02 |
| | 15 | 4.7 ± 0.7 | 0.32 ± 0.03 |
| | 16 | 5.2 ± 0.8 | 0.61 ± 0.06 |
| | 17 | 4.5 ± 0.7 | 0.67 ± 0.07 |
| | 18 | 4.9 ± 0.9 | 0.87 ± 0.09 |
| | 19 | 5.1 ± 0.9 | 1.0 ± 0.1 |
| | 20 | 5.5 ± 0.8 | 1.1 ± 0.1 |
| | 21 | 5.7 ± 0.8 | 1.2 ± 0.1 |
| | 22 | 5.7 ± 0.8 | 1.2 ± 0.1 |
| | 23 | 5.6 ± 0.8 | 1.2 ± 0.1 |
| $^{122}$Te$(\gamma, n)^{121m}$Te | 13 | 0.32 ± 0.05 | 0.006 ± 0.001 |
| | 14 | 0.40 ± 0.06 | 0.015 ± 0.002 |
| | 15 | 0.55 ± 0.08 | 0.034 ± 0.004 |
| | 16 | 0.7 ± 0.1 | 0.08 ± 0.01 |
| | 17 | 0.7 ± 0.1 | 0.10 ± 0.01 |
| | 18 | 0.8 ± 0.1 | 0.15 ± 0.02 |
| | 19 | 0.9 ± 0.1 | 0.17 ± 0.02 |
| | 20 | 1.0 ± 0.2 | 0.20 ± 0.02 |
| | 21 | 1.1 ± 0.2 | 0.22 ± 0.02 |
| | 22 | 1.1 ± 0.2 | 0.23 ± 0.02 |
| $^{120}$Te$(\gamma, n)^{119m}$Te | 23 | 1.1 ± 0.2 | 0.23 ± 0.02 |

| Reaction | | | |
|---|---|---|---|
| $^{122}$Te($\gamma$, n)$^{121}$Te | 10 | 0.11 ± 0.02 | 0.0005 ± 0.0001 |
| | 11 | 1.6 ± 0.2 | 0.013 ± 0.002 |
| | 12 | 3.0 ± 0.5 | 0.030 ± 0.003 |
| | 13 | 5.8 ± 1.1 | 0.12 ± 0.02 |
| | 14 | 4.8 ± 0.9 | 0.18 ± 0.03 |
| | 15 | 5.2 ± 0.9 | 0.36 ± 0.05 |
| | 16 | 5.9 ± 0.9 | 0.7 ± 0.1 |
| | 17 | 5.2 ± 0.9 | 0.8 ± 0.1 |
| | 18 | 5.8 ± 1.0 | 1.0 ± 0.2 |
| | 19 | 6.0 ± 1.1 | 1.1 ± 0.2 |
| | 20 | 6.5 ± 1.1 | 1.3 ± 0.2 |
| | 21 | 6.8 ± 1.2 | 1.4 ± 0.2 |
| | 22 | 6.8 ± 1.2 | 1.4 ± 0.2 |
| | 23 | 6.7 ± 1.2 | 1.4 ± 0.2 |
| $^{123}$Te($\gamma$, $\gamma$`)$^{123m}$Te+ $^{124}$Te($\gamma$, n)$^{123m}$Te+ $^{125}$Te($\gamma$, 2n)$^{123m}$Te | 10 | 0.43 ± 0.07 | 0.0020 ± 0.0002 |
| | 11 | 0.27 ± 0.04 | 0.0023 ± 0.0002 |
| | 12 | 0.41 ± 0.06 | 0.0041 ± 0.0005 |
| | 13 | 0.9 ± 0.1 | 0.019 ± 0.002 |
| | 14 | 1.0 ± 0.2 | 0.038 ± 0.004 |
| | 15 | 1.2 ± 0.2 | 0.08 ± 0.01 |
| | 16 | 1.6 ± 0.2 | 0.19 ± 0.02 |
| | 17 | 1.6 ± 0.2 | 0.23 ± 0.02 |
| | 18 | 1.9 ± 0.3 | 0.34 ± 0.03 |
| | 19 | 2.0 ± 0.3 | 0.38 ± 0.04 |
| | 20 | 2.3 ± 0.3 | 0.46 ± 0.05 |
| | 21 | 2.4 ± 0.4 | 0.50 ± 0.05 |
| | 22 | 2.5 ± 0.4 | 0.54 ± 0.06 |
| | 23 | 2.6 ± 0.4 | 0.56 ± 0.06 |
| $^{125}$Te($\gamma$, $\gamma$`)$^{125m}$Te+ $^{126}$Te($\gamma$, n)$^{125m}$Te | 10 | 9 ± 2 | 0.04 ± 0.01 |
| | 11 | 9 ± 2 | 0.07 ± 0.02 |
| | 12 | 10 ± 3 | 0.13 ± 0.03 |
| | 13 | 11 ± 3 | 0.21 ± 0.05 |
| | 14 | 11 ± 2 | 0.41 ± 0.08 |
| | 15 | 11 ± 2 | 0.8 ± 0.2 |
| | 16 | 11 ± 3 | 1.3 ± 0.3 |
| | 17 | 12 ± 2 | 1.7 ± 0.3 |
| | 18 | 13 ± 3 | 2.4 ± 0.4 |
| | 19 | 13 ± 3 | 2.5 ± 0.4 |
| | 20 | 13 ± 3 | 2.7 ± 0.5 |
| | 21 | 14 ± 3 | 2.9 ± 0.5 |
| | 22 | 14 ± 3 | 3.0 ± 0.4 |
| | 23 | 15 ± 3 | 3.1 ± 0.5 |
| $^{128}$Te($\gamma$, n)$^{127}$Te | 10 | 41 ± 6 | 0.19 ± 0.02 |
| | 11 | 57 ± 7 | 0.47 ± 0.05 |
| | 12 | 62 ± 7 | 0.59 ± 0.07 |
| | 13 | 56 ± 7 | 1.1 ± 0.2 |
| | 14 | 60 ± 7 | 2.2 ± 0.3 |
| | 15 | 58 ± 8 | 3.9 ± 0.4 |
| | 16 | 58 ± 8 | 6.6 ± 0.7 |
| | 17 | 53 ± 8 | 7.6 ± 0.9 |
| | 18 | 58 ± 8 | 10 ± 1 |
| | 19 | 56 ± 9 | 11 ± 1 |
| | 20 | 57 ± 9 | 12 ± 1 |
| | 21 | 57 ± 9 | 12 ± 1 |
| | 22 | 56 ± 9 | 12 ± 1 |
| | 23 | 56 ± 9 | 12 ± 1 |
| $^{130}$Te($\gamma$, $n$)$^{129g}$Te | 10 | 100 ± 15 | 0.47 ± 0.05 |
| | 11 | 98 ± 10 | 0.81 ± 0.09 |
| | 12 | 96 ± 10 | 0.95 ± 0.09 |
| | 13 | 89 ± 10 | 1.8 ± 0.2 |
| | 14 | 88 ± 10 | 3.3 ± 0.3 |
| | 15 | 85 ± 10 | 5.8 ± 0.6 |
| | 16 | 81 ± 10 | 9.4 ± 0.9 |
| | 17 | 83 ± 10 | 12 ± 1 |
| | 18 | 80 ± 11 | 14 ± 1 |
| | 19 | 80 ± 11 | 15 ± 2 |
| | 20 | 79 ± 10 | 16 ± 2 |
| | 21 | 79 ± 10 | 16 ± 2 |
| | 22 | 78 ± 10 | 17 ± 2 |
| | 23 | 79 ± 10 | 17 ± 2 |
| $^{130}$Te($\gamma$, $n$)$^{129m}$Te | 11 | 2.0 ± 0.4 | 0.019 ± 0.003 |
| | 12 | 4.0 ± 0.8 | 0.046 ± 0.007 |
| | 13 | 11 ± 2 | 0.22 ± 0.02 |
| | 14 | 12 ± 2 | 0.46 ± 0.05 |
| | 15 | 15 ± 2 | 1.0 ± 0.1 |
| | 16 | 19 ± 3 | 2.2 ± 0.2 |
| | 17 | 17 ± 3 | 2.8 ± 0.3 |
| | 18 | 20 ± 3 | 3.5 ± 0.4 |
| | 19 | 20 ± 3 | 3.9 ± 0.4 |
| | 20 | 21 ± 3 | 4.3 ± 0.4 |
| | 21 | 21 ± 3 | 4.4 ± 0.4 |
| | 22 | 22 ± 3 | 4.7 ± 0.5 |
| | 23 | 21 ± 3 | 4.7 ± 0.5 |
| $^{130}$Te($\gamma$, $n$)$^{129}$Te | 10 | | 0.47 ± 0.05 |
| | 11 | | 0.83 ± 0.09 |
| | 12 | | 1.0 ± 0.1 |
| | 13 | | 2.0 ± 0.2 |
| | 14 | | 3.8 ± 0.4 |
| | 15 | | 6.8 ± 0.7 |
| | 16 | | 11 ± 1 |
| | 17 | | 15 ± 2 |
| | 18 | | 17 ± 2 |
| | 19 | | 19 ± 2 |
| | 20 | | 20 ± 3 |
| | 21 | | 20 ± 3 |
| | 22 | | 22 ± 3 |
| | 23 | | 22 ± 3 |
| $^{123}$Te($\gamma$, $p$)$^{122}$Sb | 10 | 0.005 ± 0.001 | 0.00003 ± 0.00001 |
| | 11 | 0.007 ± 0.001 | 0.00006 ± 0.00001 |
| | 12 | 0.013 ± 0.002 | 0.00013 ± 0.00002 |
| | 13 | 0.015 ± 0.002 | 0.00030 ± 0.00004 |
| | 14 | 0.011 ± 0.002 | 0.00041 ± 0.00005 |
| | 15 | 0.013 ± 0.002 | 0.0009 ± 0.0001 |
| | 16 | 0.014 ± 0.002 | 0.0016 ± 0.0002 |
| | 17 | 0.015 ± 0.002 | 0.0021 ± 0.0002 |
| | 18 | 0.017 ± 0.003 | 0.0030 ± 0.0004 |
| | 19 | 0.019 ± 0.003 | 0.0036 ± 0.0005 |
| | 20 | 0.023 ± 0.004 | 0.005 ± 0.001 |
| | 21 | 0.028 ± 0.004 | 0.006 ± 0.001 |
| | 22 | 0.032 ± 0.005 | 0.007 ± 0.001 |
| | 23 | 0.032 ± 0.006 | 0.007 ± 0.001 |
| $^{125}$Te($\gamma$, $p$)$^{124}$Sb | 19 | 0.012 ± 0.003 | 0.0024 ± 0.0005 |
| | 20 | 0.035 ± 0.006 | 0.007 ± 0.001 |
| | 21 | 0.053 ± 0.009 | 0.011 ± 0.001 |
| | 22 | 0.08 ± 0.01 | 0.017 ± 0.002 |
| | 23 | 0.11 ± 0.02 | 0.024 ± 0.003 |
| $^{128}$Te($\gamma$, $p$)$^{127}$Sb | 17 | 0.006 ± 0.001 | 0.0009 ± 0.0001 |
| | 18 | 0.010 ± 0.002 | 0.0018 ± 0.0003 |
| | 19 | 0.022 ± 0.005 | 0.004 ± 0.001 |
| | 20 | 0.032 ± 0.005 | 0.006 ± 0.001 |
| | 21 | 0.07 ± 0.01 | 0.014 ± 0.002 |
| | 22 | 0.15 ± 0.03 | 0.032 ± 0.004 |
| | 23 | 0.22 ± 0.04 | 0.048 ± 0.005 |
| $^{130}$Te($\gamma$, $p$)$^{129}$Sb | 19 | 0.007 ± 0.001 | 0.0014 ± 0.0002 |
| | 20 | 0.013 ± 0.002 | 0.0027 ± 0.0004 |
| | 21 | 0.020 ± 0.003 | 0.0042 ± 0.0005 |
| | 22 | 0.07 ± 0.01 | 0.014 ± 0.002 |
| | 23 | 0.11 ± 0.02 | 0.025 ± 0.003 |

## APPENDIX 2: TALYS PARAMETERS AND ACCOUNTING OF ISOSPIN SPLITTING IN CMPR

The essential components of TALYS calculations for photonuclear reaction cross sections are the nuclear level densities and γ-ray strength functions. The cross sections of the reactions are computed in this work using TALYS 2.0 with standard parameters. For the reactions $^{114}$Cd($\gamma$,$p$)$^{113}$Ag and $^{128}$Te($\gamma$,$p$)$^{127}$Sb, the effects of altering a number of input options are examined, including level densities (Constant Temperature + Fermi gas model, Back-shifted Fermi gas Model, Generalised Superfluid Model, Skyrme-Hartree-Fock-Bogolyubov level densities from numerical tables, Skyrme-Hartree-Fock-Bogolyubov combinatorial level densities from numerical tables, Temperature-dependent Gogny-Hartree-Fock-Bogolyubov combinatorial level densities from numerical tables) and γ-strength functions (Kopecky-Uhl generalized Lorentzian, Brink-Axel Lorentzian, Hartree-Fock BCS tables, Hartree-Fock-Bogoliubov tables, Goriely's hybrid model, Goriely T-dependent HFB, T-dependent RMF, Gogny D1M HFB+QRPA, Simplified Modified Lorentzian Model, Skyrme HFB+QRPA).

Using these options, the cross section per equivalent quantum for the reactions $^{114}$Cd($\gamma$,$p$)$^{113}$Ag and $^{128}$Te($\gamma$,$p$)$^{127}$Sb are displayed in Figures 1 and 2. Changing the parameters LD 1–6 and GSF 1–10 leads to 60 different results for theoretical $\sigma(E)$ and $\sigma_q^{nat}(E)$. As seen in the Fig. 1 and Fig. 2, the change of these input options has no impact on the TALYS results. Thus, it is confirmed that the difference between the TALYS and CMPR results is due to the consideration of isospin splitting in CMPR. A brief description of isospin splitting is given below.

In nuclei with $N \neq Z$, upon absorption of electric dipole γ photons, two branches of the GDR are excited, $T_< = T_0$ and $T_> = T_0 + 1$, where $T_0 = \frac{|N-Z|}{2}$ [49]. Fig. 3 shows the excitations of the isospin components $T_<$ and $T_>$ of the GDR in initial nucleus $(N, Z)$ and their decay according to the proton $(N, Z - 1)$ and neutron $(N - 1, Z)$ channels. From Fig. 3, it can be observed that the decay of excited GDR states with isospin $T_> = T_0 + 1$ according to the neutron channel to low-lying states $T = T_0 - 1/2$ with neutron emission is forbidden, which leads to an increase in the reaction cross section (γ,p) and to a maximum shift of the reaction cross section (γ,p) with respect to reactions (γ,n) towards higher energies in the nucleus $(N, Z)$.

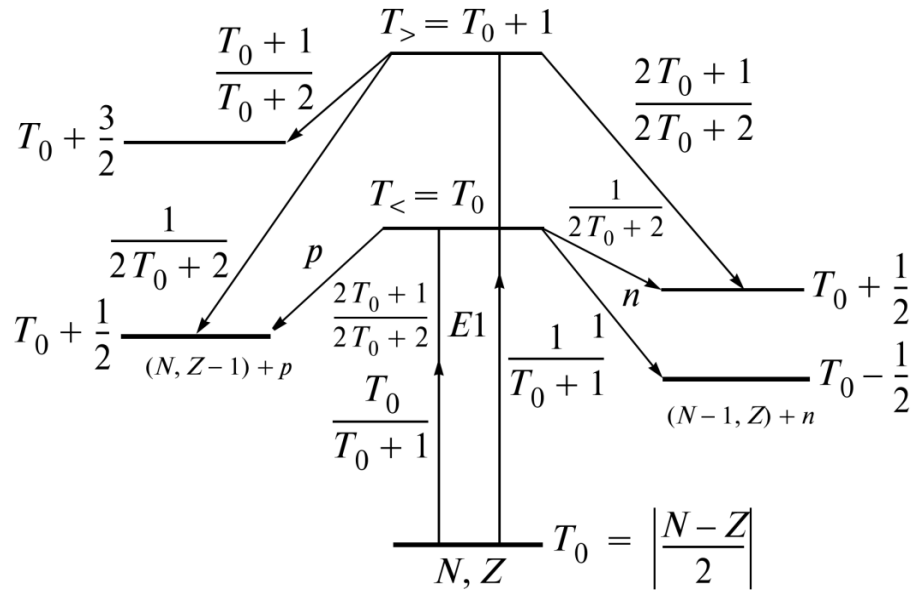

Figure 3. Scheme of excitation of states $T_<$ and $T_>$ in the nucleus $(N, Z)$ and their decay along the proton channel $(N, Z - 1)$ and neutron channel $(N - 1, Z)$

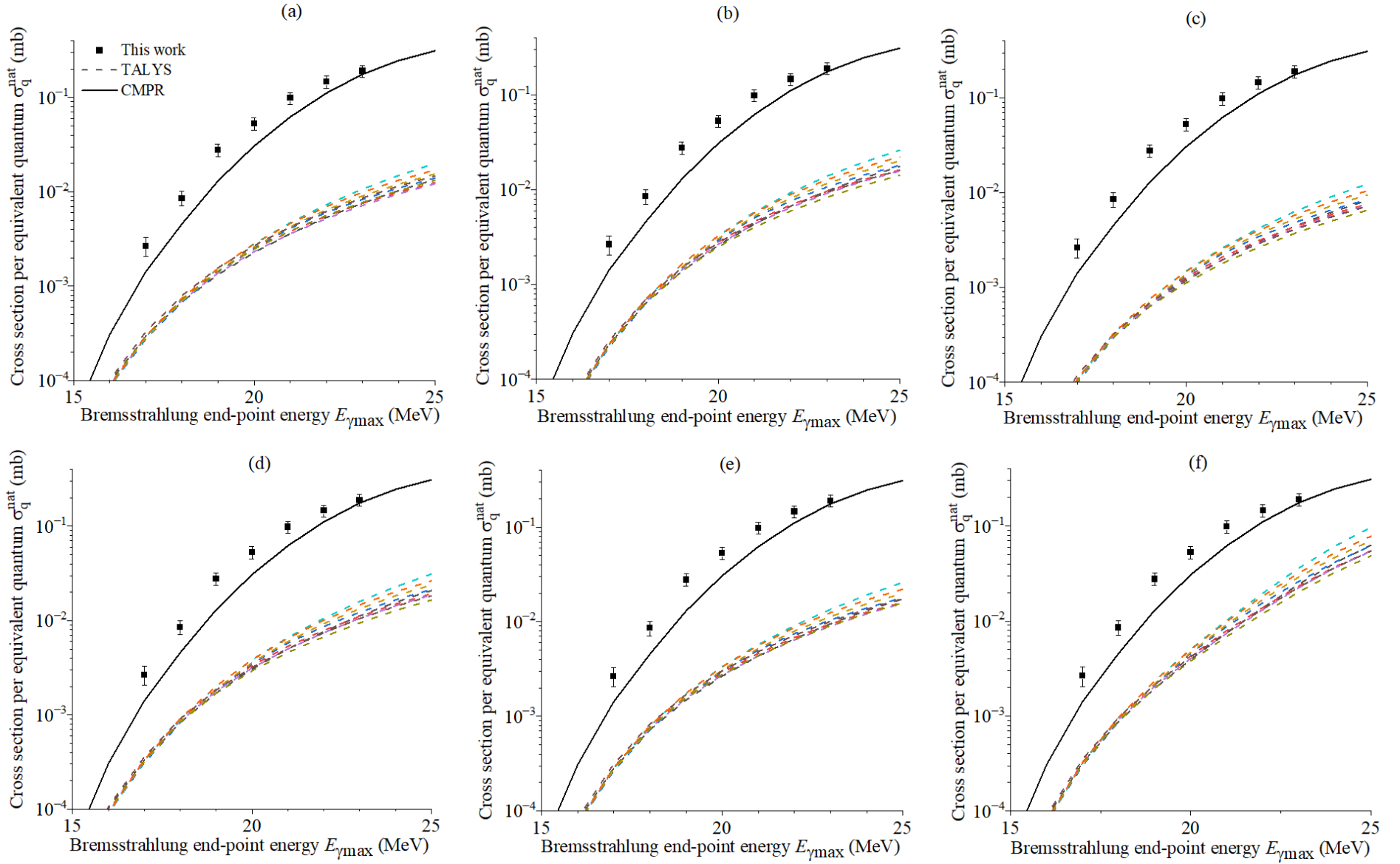

Figure 1. Cross sections per equivalent quantum $\sigma_q^{nat}(E)$ for the $^{114}$Cd$(\gamma,p)^{113}$Ag reaction calculated with the TALYS code for six level density models $LD1$-$LD6$ (a-f) and ten gamma strength functions (dashed lines) as well as simulated values using the CMPR (solid lines)

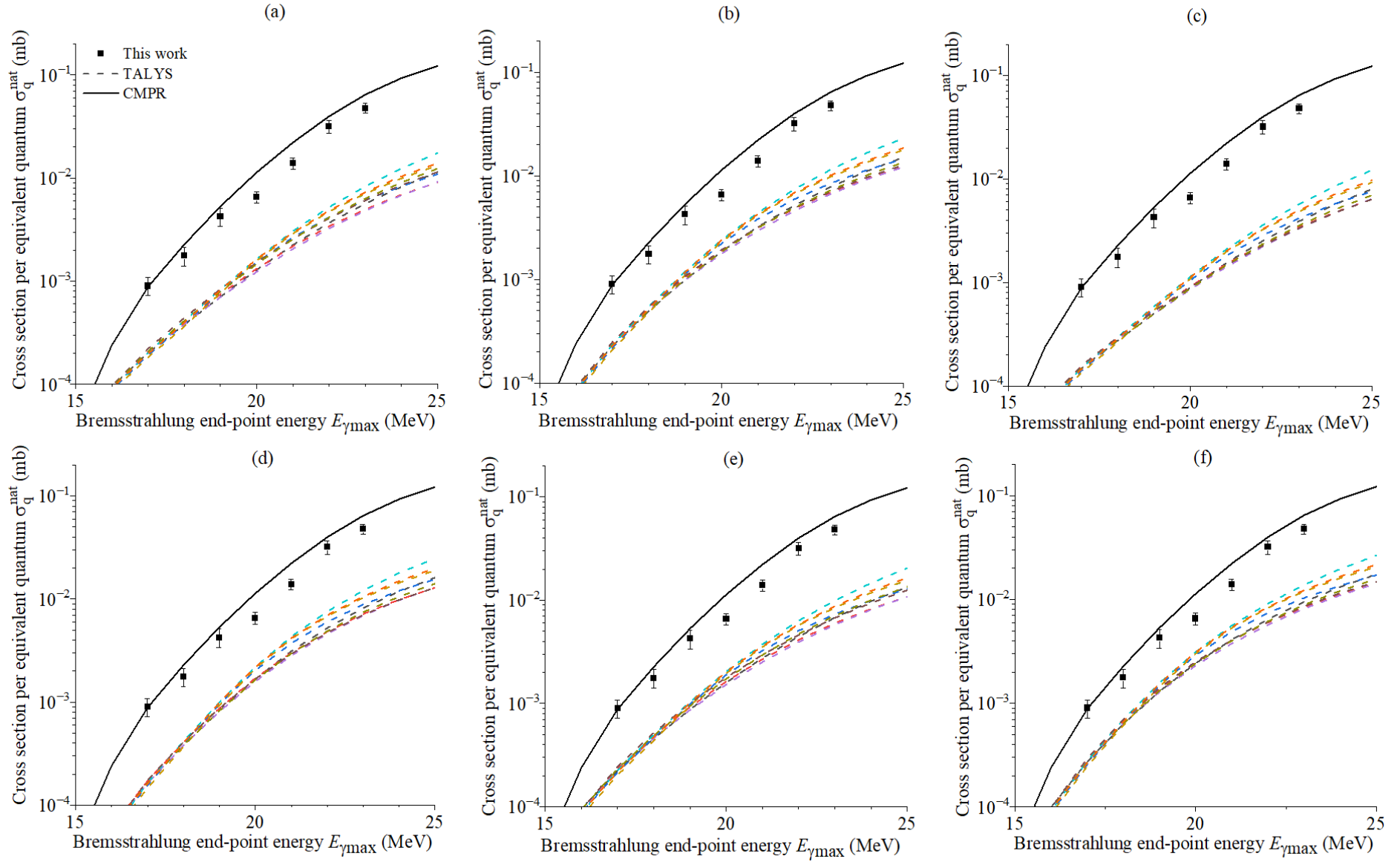

Figure 2. Cross sections per equivalent quantum $\sigma_q^{nat}(E)$ for the $^{128}$Te$(\gamma,p)^{127}$Sb reaction calculated with the TALYS code for six level density models $LD1$-$LD6$ (a-f) and ten gamma strength functions (dashed lines) as well as simulated values using the CMPR (solid lines)

The value of isospin splitting of the GDR is determined by the following relation [50]:

$$\Delta E = E(T_>) - E(T_<) = \frac{60}{A}(T_0 + 1) \quad (1)$$

The ratio of the probabilities of excitation of states $T_>$ and $T_<$ is described by the following relation [51]:

$$\frac{\sigma(T_>)}{\sigma(T_<)} = \frac{1}{T_0} \frac{1 - 1.5 T_0 A^{-2/3}}{1 + 1.5 A^{-2/3}} \quad (2)$$

Table 1 shows the values of the energy of the GDR isospin energy splitting calculated on the basis of relations (1) for isotopes $^{nat}Cd$ and $^{nat}Te$. Also Table 1 contains integral cross sections $\sigma_<^{int}$ and $\sigma_>^{int}$ of the isospin components reactions ($\gamma$, s$n$) = ($\gamma$, $n$) + ($\gamma$, $np$) + ($\gamma$, $2n$) and ($\gamma$, s$p$) = ($\gamma$, $p$) + ($\gamma$, $np$) + ($\gamma$, $2p$) in the energy region below 40 MeV, the ratio of the reaction cross sections $\frac{\sigma_>^{int}}{\sigma_<^{int}}$, calculated on the basis of CMPR for isotopes $^{nat}Cd$ and $^{nat}Te$. Table 1 shows that, for $^{nat}Cd$, the growth of the mass number A from 106 to 116 results in increase of the isospin energy splitting by the value ≈2.29 MeV. For $^{nat}Te$, the growth of A from 120 to 130, leads to the increase of energy splitting by the value ≈2.16 MeV. The isospin splitting leads to the shift of the proton cross section of the relatively neutron in the side of the high energy.

Table 1. CMPR results for GDR isospin energy splitting,, integral cross sections $\sigma_<^{int}$ and $\sigma_>^{int}$ of the isospin components reactions ($\gamma$, s$n$) and ($\gamma$, s$p$), the ratio of the reaction cross sections $\frac{\sigma_>^{int}}{\sigma_<^{int}}$, calculated on the basis of CMPR

|  | | | | | | | | |
|---|---|---|---|---|---|---|---|---|
| $^{nat}Cd$ | | | | | | | | |
| | | | ($\gamma$, s$n$) | | | ($\gamma$, s$p$) | | |
| A | $T_0$ | $E(T_>)-E(T_<)$ | $\sigma_<^{int}$ (MeVmb) | $\sigma_>^{int}$ (MeVmb) | $\frac{\sigma_>^{int}}{\sigma_<^{int}}$ | $\sigma_<^{int}$ (MeVmb) | $\sigma_>^{int}$ (MeVmb) | $\frac{\sigma_>^{int}}{\sigma_<^{int}}$ |
| 106 | 5 | 3.40 | 1405 | 42 | 0.03 | 193 | 210 | 1.09 |
| 108 | 6 | 3.89 | 1560 | 42 | 0.03 | 92 | 156 | 1.70 |
| 110 | 7 | 4.36 | 1648 | 45 | 0.03 | 58 | 112 | 1.92 |
| 111 | 7.5 | 4.59 | 1716 | 66 | 0.04 | 57 | 99 | 1.73 |
| 112 | 8 | 4.82 | 1711 | 49 | 0.03 | 47 | 77 | 1.64 |
| 113 | 8.5 | 5.04 | 1778 | 69 | 0.04 | 46 | 60 | 1.30 |
| 114 | 9 | 5.26 | 1772 | 50 | 0.03 | 41 | 48 | 1.17 |
| 116 | 10 | 5.69 | 1823 | 46 | 0.03 | 36 | 27 | 0.75 |
| $^{nat}Te$ | | | | | | | | |
| | | | ($\gamma$, s$n$) | | | ($\gamma$, s$p$) | | |
| A | $T_0$ | $E(T_>)-E(T_<)$ | $\sigma_<^{int}$ (MeVmb) | $\sigma_>^{int}$ (MeVmb) | $\frac{\sigma_>^{int}}{\sigma_<^{int}}$ | $\sigma_<^{int}$ (MeVmb) | $\sigma_>^{int}$ (MeVmb) | $\frac{\sigma_>^{int}}{\sigma_<^{int}}$ |
| 120 | 8 | 4.50 | 1819 | 43 | 0.02 | 63 | 108 | 1.71 |
| 122 | 9 | 4.92 | 1891 | 46 | 0.02 | 40 | 74 | 1.85 |
| 123 | 9.5 | 5.12 | 1964 | 63 | 0.03 | 39 | 62 | 1.59 |
| 124 | 10 | 5.32 | 1943 | 47 | 0.02 | 34 | 47 | 1.38 |
| 125 | 10.5 | 5.52 | 2015 | 49 | 0.02 | 35 | 49 | 1.40 |
| 126 | 11 | 5.71 | 1995 | 42 | 0.02 | 30 | 31 | 1.03 |
| 128 | 12 | 6.09 | 2043 | 36 | 0.02 | 28 | 18 | 0.64 |
| 130 | 13 | 6.46 | 2084 | 28 | 0.01 | 27 | 12 | 0.44 |

**Acknowledgment**

The authors would like to thank the staff of the MT-25 microtron of the Flerov Laboratory of Nuclear Reactions, Joint Institute for Nuclear Research, for their cooperation in the realization of the experiments. The work was supported by the National Research Foundation of South Africa in association with the Joint Institute for Nuclear Research (project № 22 Radiochemistry (FLNR part)) in part.

Financial support: The work is financed by the state budget of the Republic of Uzbekistan.